\documentclass[preprint,3p,10pt]{elsarticle} 

\usepackage{siunitx}
\usepackage[export]{adjustbox}

\usepackage{graphicx}
\usepackage{units}

\usepackage{booktabs}

\usepackage{multirow}

\usepackage{pgfplots}
\usepackage{pgfplotstable}
\usepackage{tikz}
\usepackage{environ}
\pgfplotsset{compat=newest}
\pgfplotsset{plot coordinates/math parser=false}
\usepgfplotslibrary{groupplots}
\usetikzlibrary{calc,shapes,patterns,backgrounds,arrows,positioning,pgfplots.groupplots,matrix,decorations.pathmorphing,fit}

\usepgfplotslibrary{external}
\usepgfplotslibrary{groupplots}
\makeatletter
\newsavebox{\measure@tikzpicture}
\NewEnviron{scaletikzpicturetowidth}[1]{%
	\def\tikz@width{#1}%
	\begin{lrbox}{\measure@tikzpicture}%
		\BODY
	\end{lrbox}%
	\pgfmathparse{#1/\wd\measure@tikzpicture}%
	\BODY
}
\makeatother

\usepackage{amsmath,amssymb}
\usepackage{xcolor}

\definecolor{redVLE}{RGB}{213,62,79}
\definecolor{blueVLE}{RGB}{50,136,189}

\definecolor{color1}{RGB}{228,26,28}
\definecolor{color2}{RGB}{55,126,184}
\definecolor{color3}{RGB}{77,175,74}
\definecolor{color4}{RGB}{152,78,163}

\definecolor{color1Pastel}{RGB}{102,194,165}
\definecolor{color2Pastel}{RGB}{252,141,98}
\definecolor{color3Pastel}{RGB}{141,160,203}
\definecolor{color4Pastel}{RGB}{231,138,195}

\definecolor{colora}{RGB}{228,26,28}
\definecolor{colorb}{RGB}{55,126,184}
\definecolor{colorc}{RGB}{77,175,74}
\definecolor{colord}{RGB}{152,78,163}
\definecolor{colore}{RGB}{255,127,0}

\definecolor{lightGray}{RGB}{217,217,217}

\newcommand{\at}[2][]{#1|_{#2}}

\pgfplotsset{select coords between index/.style 2 args={
		x filter/.code={
			\ifnum\coordindex<#1\def\pgfmathresult{}\fi
			\ifnum\coordindex>#2\def\pgfmathresult{}\fi
		}
}}

\usetikzlibrary{decorations.markings}

\makeatletter
\def\ps@pprintTitle{%
	\let\@oddhead\@empty
	\let\@evenhead\@empty
	\def\@oddfoot{}%
	\let\@evenfoot\@oddfoot}
\makeatother

\usepackage{setspace}

\usepackage{natbib}

\begin{document}

	\begin{frontmatter}
		
		\vspace*{60pt}
		\title{\textbf{Thermodynamic analysis and large-eddy simulations of LOx-CH$_4$ and LOx-H$_2$ flames at high pressure\tnoteref{t1}}}
			
			\tnotetext[t1]{Preprint submitted to AIAA Joint Propulsion Conference 2018, Cincinatti, Ohio}
			
			\author[UniBw]{Christoph Traxinger\corref{cor1}}
			\cortext[cor1]{Corresponding author: christoph.traxinger@unibw.de}
			\author[UniBw]{Julian Zips}
			\author[UniBw]{Michael Pfitzner}
			\address[UniBw]{Institute for Thermodynamics, Bundeswehr University Munich,Werner-Heisenberg-Weg 39, 85577 Neubiberg, Germany}
			
			\begin{abstract}
				Under rocket-relevant conditions, real-gas effects and thermodynamic non-idealities are prominent features of the flow field. Experimental investigations indicate that phase separation can occur depending on the operating conditions and on the involved species in the multicomponent flow. During the past decades, several research groups in the rocket combustion community have addressed this topic. In this contribution we employ a high-fidelity thermodynamic framework comprising real-gas and multicomponent phase separation effects to investigate liquid oxygen-methane and liquid oxygen-hydrogen flames at high pressure. A thorough introduction and discussion on multicomponent phase separation is conducted. The model is validated with experimental data and incorporated in a reacting flow CFD code. Thermodynamic effects are presented using one-dimensional counterflow diffusion flames. Both real-gas and phase separation effects are present and quantified in terms of derived properties. Finally, the method is applied in a three-dimensional large eddy simulation of a single-element reference test case and the results are compared to experimental data. 
			\end{abstract}
			
		\end{frontmatter}
		

\section{Introduction}

Main stage liquid-propellant rocket engines (LREs) are typically operated at elevated combustion chamber pressures up to 250~bar, which is significantly higher than the critical pressure of the propellants and of most combustion products. In addition to the high pressure, the cryogenic injection temperature of the oxidizer which is usually oxygen is a main characteristic of LREs. Typical injection temperatures are ranging from 80~K to 120~K and oxygen is therefore injected as a compressed liquid. This kind of injection condition is often referred to as "trans-critical" as the oxygen is heated up subsequent to the injection and therefore undergoes a transition from an initially liquid-like (high density) to a gas-like (low density) state. This process features strong real-gas effects which imply that the thermodynamic properties are dependent on both temperature as well as pressure. The result of this dependency is a non-linear behavior of different thermodynamic properties. Additional non-linearities arise from the multicomponent mixtures which are caused by the combustion process.\\

In the field of experimental investigations Mayer \textit{et al.}~\cite{Mayer1996,mayer1998a} performed pioneering work. They investigated pressure effects on injection and mixing of both reacting and non-reacting jets. For pure fluid injection into itself Mayer \textit{et al.}~\cite{mayer1998a} demonstrated that at supercritical pressure phase boundaries and heat of vaporization vanish. Therefore, the sharp interface is replaced by a diffuse one with finger-like structures and the fluid features real-gas effects. Identical findings have been reported by Chehroudi~\cite{chehroudi2012a} and Oschwald \textit{et al.}~\cite{oschwald2006a}. By filling the chamber with a different gas~\cite{mayer1998a} the jet was brought back to a subcritical behavior although the pressure was supercritical with respect to the pure fluids' critical point. As a reacting jet Mayer \textit{et al.}~\cite{mayer1998a} investigated an oxygen-hydrogen combustion case (LOx/H$_2$). By changing the pressure, they managed to create a sub- and supercritical case where phenomena similar to the inert jets were deduced. The work of Mayer \textit{et al.}~\cite{Mayer1996,mayer1998a} was a first important step for the investigation of flow phenomena and thermodynamic states occurring under rocket-relevant and engine-relevant conditions. Different groups~\cite{roy2013a,manin2014a,muthukumaran2014a,falgout2016a,crua2017a,traxinger2017a} followed their example. They all investigated inert cases and provided mainly visual information on jets under elevated pressure conditions covering both single- as well as multi-phase phenomena. A key message from all these groups concerning the transition between single- and multi-phase phenomena is that the interaction between the injectant and the surrounding is decisive for the onset of the respective phenomena.\\

Over the past 20 years many groups have focused on the development of numerical tools in order to investigate injection and combustion under rocket-relevant conditions. Almost all apply a thermodynamic closure based on the cubic equation of state and assume a single-phase state. Pioneering work in the field of inert injection has been done by Oefelein and Yang~\cite{Oefelein1998} as well as Zong \textit{et al.}~\cite{Zong2004} who both applied Large-Eddy simulations (LESs) within their studies. Other groups\cite{schmitt2009a,kim2011a,muller2016b,ries2017a} followed their example and developed similar CFD codes. First outstanding numerical investigations of reacting flows under LRE conditions was done by Oefelein~\cite{Oefelein2005}. In his work, Oefelein~\cite{Oefelein2005} characterized a supercritical LOx/H$_2$ flame emanating from a shear-coaxial injector using direct numerical simulation and LES. Another approach to study non-premixed reacting flows at representative operating conditions is the analysis of one-dimensional counterflow diffusion flames. Amongst others, Ribert \textit{et al.}~\cite{Ribert2008}, Lacaze and Oefelein~\cite{Lacaze2012} and Banuti \textit{et al.}~\cite{banuti2016a} conducted detailed investigations of LOx/H$_2$ flames: Ribert \textit{et al.}~\cite{Ribert2008} focused on the dependency of the flame thickness and the heat release on pressure and strain rate in physical space and quantified the influence of Soret and Dufour effects. Lacaze and Oefelein~\cite{Lacaze2012} performed a detailed analysis of strain effects, pressure and temperature boundary conditions as well as real-fluid effects on the flame structure in both physical and mixture fraction space to develop a tabulated combustion model. Banuti \textit{et al.}~\cite{banuti2016a} investigated the transition of the thermodynamic state of LOx from a liquid-like state to a gas-like state to an ideal-gas state and concluded that the real-gas effects are mainly limited to the pure oxygen and thus mixing occurs under ideal gas conditions.

With methane getting more attention as future LRE fuel, similar studies have been performed on one-dimensional LOx/CH$_4$ counterflow diffusion flames. Kim \textit{et al.}~\cite{Kim2013} solve the real-gas flamelet equations in the mixture fraction space to establish a library of the thermo-chemical state incorporating oxygen/methane chemistry and non-ideal thermodynamics. Lapenna \textit{et al.}~\cite{lapenna2017a} analyze unsteady non-premixed LOx/CH$_4$ flamelets at supercritical pressures in order to study the effects of pressure and strain on autoignition, re-ignition and quenching. Slightly out of the scope of the LRE conditions,  Juan\'{o}s and Sirignano~\cite{Juanos2017} use steady air/methane as well as air/water-vapor/methane flamelets between 1 and 100~bar to investigate the pressure effects of these propellants and the required modeling.

In accordance to the described experimental investigations some of these numerical groups performed a-posteriori analysis of possible phase separation phenomena~\cite{Lacaze2012,banuti2016a,lapenna2017a}. All of them came to the finding that multi-phase effects do not occur and the mixture stays always in a single-phase state. A deficiency of these studies was that these groups applied a rapid estimation method which is known to produce sever deviations~\cite{poling2001} in complex mixtures like they occur during the combustion process. Recently, Banuti \textit{et al.}~\cite{banuti2017a} used a high fidelity method a-posteriori and pointed out that phase separation might be possible under rocket-relevant conditions but further investigations are necessary.\\

In the present study we conduct a detailed numerical analysis of both LOx/H$_2$ and LOx/CH$_4$ at representative conditions taken from two well-known experiments. Deviations from the ideal-gas state due to pressure, temperature and non-ideal mixing are considered. The analysis is based on a cubic equation of state and takes into account real-gas and possible phase-separation effects. After performing fundamental investigations using one-dimensional counterflow diffusion flames, the results of the methane flame are tabulated and applied in a three-dimensional Large-Eddy Simulation. The objective of this study is to investigate possible phase separation inside the flame and to examine its effects on the result. A consistent vapor-liquid equilibrium model~\cite{qiu2015a,matheis2018a,traxinger2017a,traxinger2018a} will be applied to account for phase separation effects.

\section{Theoretical Background and Models} \label{sec: Theoretical Background and Models}

\subsection{Thermodynamic Model}

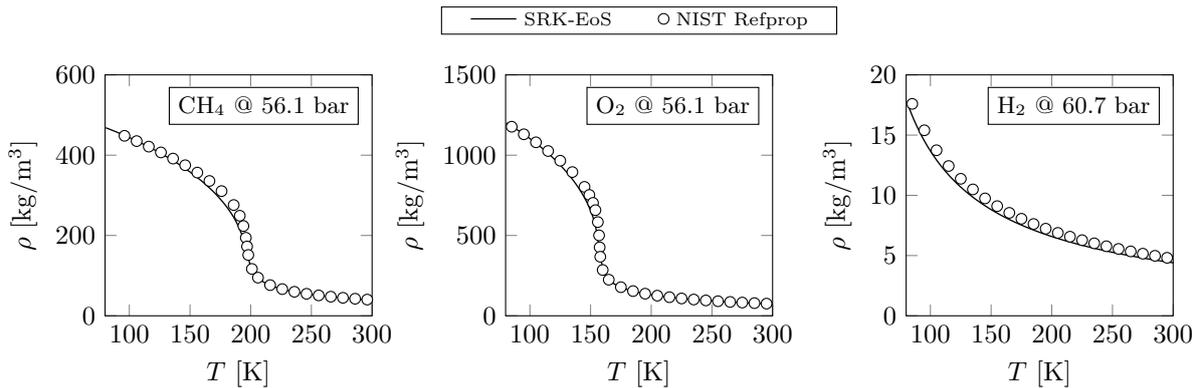
\begin{figure}[h]
	\centering
	\tikzsetnextfilename{evalEoS}%

\begin{tikzpicture}
	\begin{groupplot}[group style = {group name = plots,group size = 3 by 1, horizontal sep = 50pt}, width = 0.32\textwidth, height = 0.3\textwidth]
		\nextgroupplot
		[
		xlabel = $T\text{ [K]}$,
		xmin=80,
		xmax = 300,
		ylabel = $\rho\text{ [kg/m$^3$]}$,
		ymin=0,
		ymax = 600
		]
		\addplot [solid,line width=0.5pt] table [x=T,y=rho, col sep=comma]{../sources/Reinstoff/CH4/SoaveRedlichKwong/postProcessing/sampleDict/1e-07/line_0_T_rho.csv};
		\label{plot:SRK-EoS}
		\addplot [only marks,mark=*,fill=white] table [x=T,y=rho]{../sources/NIST/Methane_5.61MPa.dat}; \label{plot:NIST}
		\draw [-] (rel axis cs:0.95,0.95) node[anchor=north east,font=\small,draw] {CH$_4$ @ 56.1 bar};
		\coordinate (top) at (rel axis cs:0,1);
		\nextgroupplot
		[
		xlabel = $T\text{ [K]}$,
		xmin=80,
		xmax = 300,
		ylabel = $\rho\text{ [kg/m$^3$]}$,
		ymin=0,
		ymax = 1500,
		/pgf/number format/1000 sep={},
		]
		\addplot [solid,line width=0.5pt] table [x=T,y=rho, col sep=comma]{../sources/Reinstoff/O2_56.1bar/SoaveRedlichKwong/postProcessing/sampleDict/1e-07/line_0_T_rho.csv};
		\addplot [only marks,mark=*,fill=white] table [x=T,y=rho]{../sources/NIST/Oxygen_5.61MPa.dat};
		\draw [-] (rel axis cs:0.95,0.95) node[anchor=north east,font=\small,draw] {O$_2$ @ 56.1 bar};
		\nextgroupplot
		[
		xlabel = $T\text{ [K]}$,
		xmin=80,
		xmax = 300,
		ylabel = $\rho\text{ [kg/m$^3$]}$,
		ymin=0,
		ymax = 20
		]
		\addplot [solid,line width=0.5pt] table [x=T,y=rho, col sep=comma]{../sources/Reinstoff/H2/SoaveRedlichKwong/postProcessing/sampleDict/1e-07/line_0_T_rho.csv};
		\addplot [only marks,mark=*,fill=white] table [x=T,y=rho]{../sources/NIST/Hydrogen_6.07MPa.dat};	
		\draw [-] (rel axis cs:0.95,0.95) node[anchor=north east,font=\small,draw] {H$_2$ @ 60.7 bar};
		\coordinate (bot) at (rel axis cs:1,0);
	\end{groupplot}	
   	\path (top|-current bounding box.north)--
    coordinate(legendpos)
    (bot|-current bounding box.north);
   	\node[
	anchor=south,
	inner sep=0.2em,
	font=\scriptsize,
	draw
	]at([xshift=0ex,yshift=2ex]legendpos)
	{
	\hspace*{6pt} \ref{plot:SRK-EoS} SRK-EoS \hspace*{12pt} \ref{plot:NIST} NIST Refprop \hspace*{6pt}
	};
\end{tikzpicture}%

	\caption{Comparison of the SRK-EoS~\cite{soave1972equilibrium} with reference data taken from NIST Refprop~\cite{NIST} for three different fluids (methane CH$_4$, oxygen O$_2$ and hydrogen H$_2$).}
	\label{fig:evalEoS}
\end{figure}

At supercritical pressures ($p > p_c$) and rocket engine-relevant injection temperatures (cold/cryogenic oxygen and moderate fuel temperatures, respectively) the fluid's thermo-physical properties deviate from the well-known ideal-gas law and real-gas effects have to be taken into account to accurately describe the injection and combustion process~\cite{chehroudi2012a}. 
In the field of LES under rocket-relevant conditions cubic equations of state (EoS) are commonly used\cite{Oefelein2005,Schmitt2011,banuti2016a,Ribert2008,Lacaze2012,Kim2013,lapenna2017a,Juanos2017} to describe the $pvT$-behavior of real fluids. The cubic EoS account for the most important interactions, namely the intermolecular attractive forces and the co-volume of the molecules, and can be written in the following general, pressure-explicit form\cite{poling2001}:
\begin{equation}
p = \frac{\mathcal{R} \; T}{v - b} - \frac{a \left(T\right)}{v^2 + ubv + wb^2} = \frac{\mathcal{R} \; T}{v - b} - \frac{a_c \; \alpha\left(T\right)}{v^2 + ubv + wb^2} \; .
\label{eq:cubicEoS}
\end{equation}
In this equation, $\mathcal{R}$ is the gas constant, $T$ is the temperature and $v = 1/\rho$ is the specific volume. The intermolecular attractive forces and the co-volume of the molecules are considered through the parameters $a = a_c \; \alpha\left(T\right)$ and $b$, respectively. The parameters $u$ and $w$ are model constants. In the present study, methane and hydrogen are considered as fuels and oxygen as oxidizer. For these fluids the EoS derived by Soave, Redlich and Kwong~\cite{soave1972equilibrium} (SRK-EoS) is a good choice\cite{Ribert2008,Juanos2017,Kim2013} to describe the trans-critical behavior. To recover the SRK-EoS from Eq.~\eqref{eq:cubicEoS} the model constants have to be set to $u = 1$ and $w = 0$, respectively. The calculation of the other parameters is listed in Tab.~\ref{tab:cubicEoS} and can be derived from the corresponding states principle\cite{poling2001}. In Fig.~\ref{fig:evalEoS} the SRK-EoS is compared to reference data obtained from NIST Refprop~\cite{NIST}. For all three fluids a very good agreement between prediction and reference data at the relevant pressure and temperature range is found.
\begin{table}[b]
	\centering
	\caption{Parameters for Eq.~\eqref{eq:cubicEoS} to recover the SRK-EoS~\cite{soave1972equilibrium}.}
	\begin{tabular}{c|c|c|c|c|c}
		\toprule
		$u$ & $w$ & $a_c$ & $\alpha \left( T \right)$ & $\kappa$ & $b$\\
		\midrule
		1 & 0  & $0.42748 \; \frac{\mathcal{R}^2 T_c^2}{p_c}$ & $\left[ 1 + \kappa \left( 1 - \sqrt{\frac{T}{T_c}}\right)\right]^2$ & $0.480 + 1.574 \; \omega - 0.176 \; \omega^2$ & $0.08664 \; \frac{\mathcal{R} T_c}{p_c}$\\
		\bottomrule 
	\end{tabular}
	\label{tab:cubicEoS}
\end{table}

For treating multicomponent mixtures, the concept of the one-fluid mixture in combination with mixing rules\cite{poling2001} is applied:
\begin{equation}
a = \sum_{i}^{N_c}\sum_{j}^{N_c} \xi_i \xi_j a_{ij} \hspace{1cm} \text{and} \hspace{1cm} b = \sum_{i}^{N_c} \xi_i b_i \; .
\label{eq:ab_PR-EoS}
\end{equation}
Here, $\xi_i$ is the mole fraction of species $i$, whereby in the following we denote the overall mole fraction by $\mathbf{z} = \{ z_1 , ... , z_{N_c} \}$ and the liquid and vapor mole fractions by $\mathbf{x} = \{ x_1 , ... , x_{N_c} \}$ and $\mathbf{y} = \{ y_1 , ... , y_{N_c} \}$, respectively. The diagonal elements of $a_{ij}$ ($i=j$) are calculated based on the respective critical parameters of the pure components, see Tab.~\ref{tab:critical properties}. In contrast, the off-diagonal elements of $a_{ij}$ ($i \neq j$) are estimated using the pseudo-critical combination rules\cite{reid1987}:
\begin{equation}
\begin{array}{l}
\omega_{ij} = 0.5 \left( \omega_i + \omega_j \right) \; , \;\;\; v_{c,ij} = \frac{1}{8} \left( v_{c,i}^{1/3} + v_{c,j}^{1/3} \right)^3 \; , \;\;\; Z_{c,ij} = 0.5 \left( Z_{c,i} + Z_{c,j} \right) \; , \;\;\;\\
\vspace{0pt}\\
T_{c,ij} = \sqrt{T_{c,i} T_{c,j}} \left(1- k_{ij} \right) \hspace{1cm} \textrm{and} \hspace{1cm} p_{c,ij} = Z_{c,ij} \mathcal{R} T_{c,ij} /v_{c,ij} \; .\\
\vspace{-8pt}
\end{array}
\label{eq:pseudoCritical}
\end{equation}
In these equations, $k_{ij}$ is the binary interaction parameter, which can be used to fit the cubic EoS to available experimental fluid data. The necessity of this fit will be thoroughly discussed subsequent to the presentation of the complete thermodynamic model applied in this study.\\

In addition to the thermal EoS, we use the departure function formalism, see, e.g., Poling \textit{et~al.}~\cite{poling2001}, to calculate the caloric properties, like enthalpy $h$ and specific heat $c_p$. The reference condition is determined using the seven-coefficient NASA polynomials proposed by Goos \textit{et~al.}~\cite{NASA}. Empirical correlations by Chung \textit{et~al.}~\cite{chung1988a} are used to estimate the viscosity and the thermal conductivity. The diffusion coefficient is deduced from the thermal conductivity under the assumption of a unity Lewis number~\cite{Schmitt2011}.\\

\begin{table}[h]
	\centering
	\caption{\label{tab:critical properties} Critical properties and acentric factor $\omega$ of oxygen $\textrm{O}_2$, hydrogen $\textrm{H}_2$ and methane $\textrm{CH}_4$.}
	\begin{tabular}{ccc|ccc|ccc}
		\multicolumn{3}{c|}{O$_{2}$} 	& \multicolumn{3}{c|}{H$_{2}$} & \multicolumn{3}{c}{CH$_{4}$}	\\\hline\hline 
		$T_{c}$ [K]	&	$p_{c}$ [MPa] & $\omega$ [-] & $T_{c}$ [K]	&	$p_{c}$ [MPa] & $\omega$ [-] & $T_{c}$ [K]	&	$p_{c}$ [MPa] & $\omega$[-] \\\hline
		$154.581$		&	$5.043$	 & $0.0222$	&  	$33.0$		& 	$1.284$  & $-0.219$  &  	$190.564$		& 	$4.5992$  & $0.011$	\\
	\end{tabular}
\end{table}

\subsection{Phase Separation in Multicomponent Mixtures}

The thermodynamic framework presented in the previous section is usually referred to as "dense gas"-approach. It assumes a single-phase state and takes into account the real-gas effects which are present during the trans-critical injection and combustion process. This has been thoroughly investigated by many groups~\cite{Oefelein2005,Ribert2008,Schmitt2011,Lacaze2012,Kim2013,banuti2016a,lapenna2017a,Juanos2017}. Different experimental investigations~\cite{Mayer1996,mayer1998a,chehroudi2012a,oschwald2006a,roy2013a,manin2014a,muthukumaran2014a,falgout2016a,crua2017a,traxinger2017a} have shown that depending on the conditions and the components forming the multicomponent mixture, phase separation might occur under rocket-relevant conditions. Over the past 20 years, different numerical research groups from the rocket community~\cite{bellan2000a,Lacaze2012,banuti2016a,banuti2017a,lapenna2017a} have come up with questions regarding possible two-phase phenomena in LREs and the topic is still under discussion and not fully understood yet. Some of these groups~\cite{Lacaze2012,banuti2016a,lapenna2017a} used rapid estimation methods to investigate possible phase separation phenomena and came to the finding that no phase separation is likely to occur under rocket-relevant conditions. Recently, Banuti \textit{et al.}~\cite{banuti2017a} applied a high fidelity thermo-physical model a-posteriori on their LES results and found that some of their scatter points lie well within the two-phase region. Therefore, they concluded that two-phase effects might occur under rocket-relevant conditions, but further investigations have to be done. In the present study, we pick up this thread. In the next section, we will give a short introduction into multi-component phase separation and related phenomena. After that, we will discuss the deficiencies of the rapid estimation method. Next, a fully consistent approach which extends the "dense gas"-approach for the phase-separation phenomena will be presented. In the final part of this section we will validate the thermodynamic framework with experimental data taken from the literature for relevant binary mixtures.

\subsubsection{Basic Introduction}

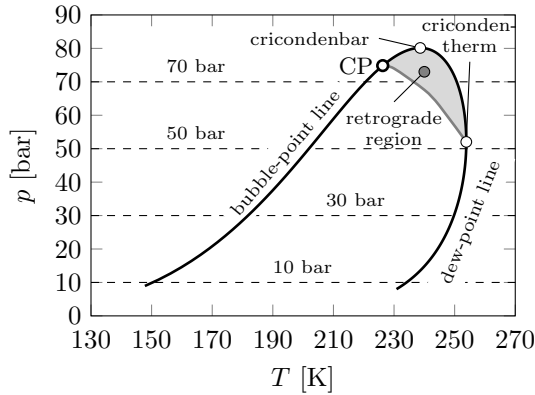
\begin{figure}[t]
	\centering
	\tikzsetnextfilename{prudhoeBayVLE}%

\begin{tikzpicture}
	\begin{axis}
	[
	width = 0.45\textwidth, 
	height = 0.35\textwidth,
	xmin=130,
	xmax=270,
	xtick = {130,150,170,190,210,230,250,270},
	xlabel = $T\text{ [K]}$,
	x tick label style={yshift=-2pt},
	ymin=0,
	ymax=90,
	ylabel = $p\text{ [bar]}$,
	ytick = {0,10,20,30,40,50,60,70,80,90},
	y filter/.code={\pgfmathparse{#1*1e-5}\pgfmathresult}
	]
	 	\draw [dashed] (axis cs: 130,70) -- (axis cs: 270,70) node [anchor=south,xshift=-4.2cm,font=\scriptsize] {70~bar};
	 	\draw [dashed] (axis cs: 130,50) -- (axis cs: 270,50) node [anchor=south,xshift=-4.2cm,font=\scriptsize] {50~bar};
	 	\draw [dashed] (axis cs: 130,30) -- (axis cs: 270,30) node [anchor=south,xshift=-2.1cm,font=\scriptsize] {30~bar};
	 	\draw [dashed] (axis cs: 130,10) -- (axis cs: 270,10) node [anchor=south,xshift=-2.8cm,font=\scriptsize] {10~bar};
	 	
	 	\node [fill=white,rotate=50,anchor=north] at (axis cs: 190,52) {\scriptsize bubble-point line};
	 	\node [fill=white,rotate=68,anchor=north] at (axis cs: 250,30) {\scriptsize dew-point line};
	 	\addplot [draw=none,fill=lightGray] table [x=T, y=p]{../sources/prudhoeBay/retrogradeRegion.dat};
	 	\addplot [color=gray,line width=1.0pt,smooth] table {
	 		226.3	74.8e5
	 		230		73.98e5
	 		240		67.60e5
	 		245		62.85e5
	 		253.84	52.03e5
	 	};
	 	
		\addplot [line width=1.0pt,forget plot] table [x=T, y=p]{../sources/prudhoeBay/pT_Diagram_PrudhoeBay.dat};
		\addplot [line width=1.0pt,forget plot,mark=*,draw=none,fill=white] table [x=T_c, y=p_c]{../sources/prudhoeBay/pc_Tc_Diagram_PrudhoeBay.dat};
		\node at (axis cs: 226.3,74.8) [anchor = east,font=\small] {CP };
		
		\draw (axis cs:238.64,80.13) -- (axis cs:225,83) node [anchor=east,font=\scriptsize] {cricondenbar};
		\draw (axis cs:253.84,52.03) -- (axis cs:255,76) node [anchor=south,font=\scriptsize,align=center] {criconden-\\therm};
		\draw (axis cs:240,73) -- (axis cs:230,64) node [anchor=north,font=\scriptsize,align=center] {retrograde\\region};
		\draw [only marks, mark=*,fill=white] plot coordinates 
		{
			(axis cs:238.64,80.13)
			(axis cs:253.84,52.03)	
		};
		\draw [only marks, mark=*,fill=gray] plot coordinates 
		{		
			(axis cs:240,73)
		};
	\end{axis}
\end{tikzpicture}%

	\caption{Phase equilibrium of the Prudhoe Bay mixture\cite{picard1987a} containing 14 components. The Peng-Robinson EoS~\cite{peng1976a} was applied for this calculation.}
	\label{fig:prudhoeBayVLE}
\end{figure}

According to classical thermodynamics~\cite{elliott2012a} a fluid can exist in three different physical states, namely solid, liquid and vapor/gas. During the injection process under typical rocket engine-relevant conditions two different states, liquid (l) and vapor (v), have been reported by experimentalists~\cite{Mayer1996,mayer1998a,chehroudi2012a,oschwald2006a}. The formation of a multi-phase mixture in an isothermal-isobaric system results from the minimization of the Gibbs energy. Hence, a phase separation occurs if and only if the Gibbs energy is lowered due to this process~\cite{michelsen2007}. Under the assumption of a thermodynamic equilibrium the resulting phase equilibrium is characterized by the equality of pressure $p$, temperature $T$ and chemical potential/specific Gibbs energy $g$ and can be expressed for a two-phase system (liquid and vapor) in the following way~\cite{firoozabadi1999}:
\begin{align}
p^\text{l} = & p^\text{v} \; , \label{eq:VLEp}\\
T^\text{l} = & T^\text{v} \; , \label{eq:VLET}\\
g_i^\text{l} = & g_i^\text{v} \; . \label{eq:VLEg}
\end{align}
This fact holds for both a pure fluid as well as a multicomponent mixture. In contrast to this, the characteristics of the critical point of a pure component
\begin{equation}
\frac{\partial p}{\partial v} \at[\bigg]{T} = 0 \hspace*{1cm} \textrm{and} \hspace*{1cm} \frac{\partial^2 p}{\partial v^2} \at[\bigg]{T} = 0
\end{equation}
are no longer applicable in multicomponent mixtures and hence the critical point (CP) has no longer to be on the top of the two-phase region. In Fig.~\ref{fig:prudhoeBayVLE} this fact can be comprehended. Here, the two-phase region of the 14 component Prudhoe Bay mixture~\cite{picard1987a} is shown. The vapor-liquid equilibrium (VLE) features some characteristics which only occur in multi-component systems. First of all, the two-phase dome does no longer collapse to a single curve (vapor-pressure curve) in the pressure temperature diagram like it is the case for pure components. Furthermore, due to the shape of the VLE two additional characteristic points arise which occur only in multicomponent systems. One is the cricondenbar point at the maximum pressure of the VLE. The other is the cricondentherm point at the maximum temperature. The latter point and the critical point enclose the retrograde region. If this region is entered via the dew-point line with an isothermal change in state, a lowering of the pressure does not result in a full condensation of the gas~\cite{kumar1987}. After the crossing of the dew-point line the vapor fraction is reduced and liquid is generated until the gray line in Fig.~\ref{fig:prudhoeBayVLE} is reached. After this point the vapor fraction increases and liquid begins to vaporize until the dew-point line is crossed again. This behavior is called retrograde condensation~\cite{kuenen1892a}. For the Prudhoe Bay mixture one can see that this can also happen at isobaric changes in state in cases where the pressure is above the corresponding critical point.

\begin{figure}[h]
	\centering
	\tikzsetnextfilename{multicomponentVLE_typeI}%
	\begin{tikzpicture}
	\begin{axis}
	[
	width = 0.40\textwidth,
	height = 0.40\textwidth,
	xmin=0,
	xmax=1,
	xlabel = {Mole fraction $x$, $y$, $z$},
	xlabel style = {rotate=-10},
	ymin=100,
	ymax=210,
	ylabel = Temperature $T$,
	ylabel style = {rotate=44},
	zmin=0.1,
	zmax=52,
	zlabel = Pressure $p$,
	x filter/.code={\pgfmathparse{1-#1}\pgfmathresult},
	z filter/.code={\pgfmathparse{#1*1e-5}\pgfmathresult},
	x dir=reverse,
	ticks=none,
	axis lines=left,
	axis line style = ultra thick,
	xlabel style = {font=\small},
	ylabel style = {font=\small},
	zlabel style = {font=\small}
	]		
		\draw (1,100,0.1) -- (1,210,0.1);
		\draw (1,210,0.1) -- (1,210,52);
		\draw (1,100,52) -- (1,210,52);
		\draw (0,210,0.1) -- (1,210,0.1);
		\draw (0,210,52) -- (1,210,52);
		\draw (0,210,0.1) -- (0,210,52);
		\addplot3 [color=black, line width=1.5pt,fill=white]  table [x=z, y=T,z=p]{../sources/criticalLocus/Methane-Nitrogen/data/VLE_Nitrogen_Methane_120K.dat};
		\addplot3 [color=black, line width=1.5pt,fill=white]  table [x=z, y=T,z=p]{../sources/criticalLocus/Methane-Nitrogen/data/VLE_Nitrogen_Methane_135K.dat};
		\addplot3 [color=black, line width=1.5pt,fill=white]  table [x=z, y=T,z=p]{../sources/criticalLocus/Methane-Nitrogen/data/VLE_Nitrogen_Methane_150K.dat};
		\addplot3 [color=black, line width=1.5pt,fill=white]  table [x=z, y=T,z=p]{../sources/criticalLocus/Methane-Nitrogen/data/VLE_Nitrogen_Methane_165K.dat};
		\addplot3 [color=black, line width=1.5pt,fill=white]  table [x=z, y=T,z=p]{../sources/criticalLocus/Methane-Nitrogen/data/VLE_Nitrogen_Methane_180K.dat};
		
		\addplot3 [color=black,dashed, line width=2.0pt]  table [x=z, y=T,z=p]{../sources/criticalLocus/Methane-Nitrogen/data/pT_Diagram_Methane_PR.dat};\label{plots:VLEvaporPressure};
		\addplot3 [color=black,dashed, line width=2.0pt]  table [x=z, y=T,z=p]{../sources/criticalLocus/Methane-Nitrogen/data/pT_Diagram_Nitrogen_PR.dat};
	
		\addplot3 [color=gray, line width=2.0pt]  table [x=zMethane, y=Tc,z=pc]{../sources/criticalLocus/Methane-Nitrogen/data/criticalLocus_MethaneNitrogen.dat};\label{plots:VLEcriticalLocus};
		\addplot3 [color=black,mark=*,only marks,line width=1.0pt,mark size = 1.5pt] table {
		0	126.192		3395800
		1	190.564		4599200
		};
		
		\draw (axis cs:0.0,190.56,45.99) -- (axis cs:0.06,193,48) node [anchor=south,xshift=-0.3ex] {CP$_\text{HV}$};
		\draw (axis cs:1.0,126.19,33.96) -- (axis cs:0.98,135,38) node [anchor=south,xshift=-0.3ex] {CP$_\text{LV}$};
		\draw (axis cs:0.65,153.6,48.1) -- (axis cs:0.65,170,51.0) node [anchor=south] {LG-Eq.};
	\end{axis}		
\end{tikzpicture}
\hspace*{70pt}
\begin{tikzpicture}
	\begin{groupplot}[group style = {group name = plots,group size = 1 by 2, vertical sep = 30pt}, width = 0.28\textwidth, height = 0.20\textwidth]
   	\nextgroupplot
    [
    xmin=0,
    xmax=1,
    xlabel = {Mole fraction $x$, $y$, $z$},
    ymin=10,
    ymax=55,
    ylabel = {Pressure $p$},
    y filter/.code={\pgfmathparse{#1*1e-5}\pgfmathresult},
    x filter/.code={\pgfmathparse{1-#1}\pgfmathresult},
    ticks=none,
    axis lines=left,
    axis line style = ultra thick,
    axis on top=true,
	xlabel style = {font=\small},
	ylabel style = {font=\small}
    ]
		\addplot [color=black, line width=1.5pt]  table [x=x, y=p]{../sources/criticalLocus/Methane-Nitrogen/data/VLE_xy_Nitrogen_Methane_150K.dat};\label{plots:VLEisotherm};
		\addplot [color=black, line width=1.5pt]  table [x=y, y=p]{../sources/criticalLocus/Methane-Nitrogen/data/VLE_xy_Nitrogen_Methane_150K.dat};
		\addplot [color=black, line width=1.5pt]  table [x=x, y=p]{../sources/criticalLocus/Methane-Nitrogen/data/VLE_xy_Nitrogen_Methane_180K.dat};
		\addplot [color=black, line width=1.5pt]  table [x=y, y=p]{../sources/criticalLocus/Methane-Nitrogen/data/VLE_xy_Nitrogen_Methane_180K.dat};
		\addplot [color=gray,line width=2.0pt,forget plot] table [x=zMethane, y=pc]{../sources/criticalLocus/Methane-Nitrogen/data/criticalLocus_MethaneNitrogen.dat};
		\addplot [color=black,mark=*,only marks,line width=1.0pt,mark size = 1.5pt] table {
		0.01	3395800
		1	4599200
		};
	\nextgroupplot
    [
    xmin=100,
    xmax=210,
    xlabel = {Temperature $T$},
    ymin=10,
    ymax=55,
    ylabel = {Pressure $p$},
    y filter/.code={\pgfmathparse{#1*1e-5}\pgfmathresult},
    ticks=none,
    axis lines=left,
    axis line style = ultra thick,
    axis on top=true,
   	xlabel style = {font=\small},
   	ylabel style = {font=\small}
    ]
		\addplot [color=black, line width=1.5pt]  table [x=T,y=p]{../sources/criticalLocus/Methane-Nitrogen/data/pT_Diagram_Nitrogen-0.4_Methane-0.6.dat};\label{plots:VLEisopleth};
		\addplot [color=black, line width=1.5pt]  table [x=T,y=p]{../sources/criticalLocus/Methane-Nitrogen/data/pT_Diagram_Nitrogen-0.6_Methane-0.4.dat};
		\addplot [color=gray,line width=2.0pt,forget plot] table [x=Tc, y=pc]{../sources/criticalLocus/Methane-Nitrogen/data/criticalLocus_MethaneNitrogen.dat};
		\addplot [color=black,line width=2.0pt,dashed,forget plot] table [x=T, y=p]{../sources/criticalLocus/Methane-Nitrogen/data/pT_Diagram_Nitrogen_PR.dat};\label{plot:vaporPressureCurve};
		\addplot [color=black,mark=*,only marks,line width=1.0pt,mark size = 1.5pt,select coords between index={0}{0},forget plot] table [x=T, y=p]{../sources/criticalLocus/Methane-Nitrogen/data/pT_Diagram_Nitrogen_PR.dat};\label{plot:criticalPoint};
		\addplot [color=black,line width=2.0pt,dashed,forget plot] table [x=T, y=p]{../sources/criticalLocus/Methane-Nitrogen/data/pT_Diagram_Methane_PR.dat};
		\addplot [color=black,mark=*,only marks,line width=1.0pt,mark size = 1.5pt,select coords between index={0}{0},forget plot] table [x=T, y=p]{../sources/criticalLocus/Methane-Nitrogen/data/pT_Diagram_Methane_PR.dat};
	\end{groupplot}
\end{tikzpicture}

%

	\caption{Principal sketch of a binary mixture with a type I critical locus.}
	\label{fig:multicomponentVLE_typeI}
\end{figure}
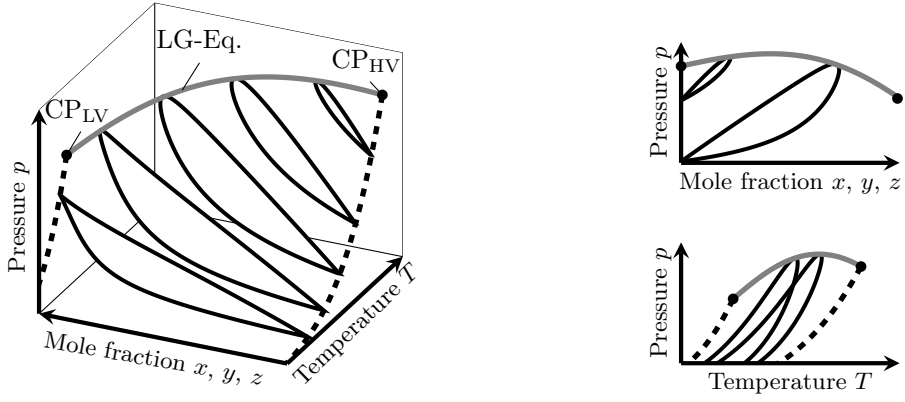

In Figure~\ref{fig:multicomponentVLE_typeI} the principal sketch of a complete binary multicomponent system with a type I critical locus~\cite{vanKonynenburg1980a} is shown. This critical locus type is the simplest one and it spans from the critical point of the low volatile (LV) component to the critical point of the high volatile (HV) fluid. Underneath this critical locus a liquid-gas equilibrium is formed which is dependent on three parameters, namely pressure $p$, temperature $T$ and overall composition $\mathbf{z}$. From this principal sketch it gets obvious that the isopleth (VLE at constant overall composition) plotted in Fig.~\ref{fig:prudhoeBayVLE} is only a single realization of the complete phase equilibrium. It is therefore mandatory to take into account the complete mixture space to assess possible phase separation phenomena and to have appropriate approaches to predict such equilibria. In the next two parts this will be discussed in more detail.

\subsubsection{Discussion of the Rapid Estimation Method}

Recently, Lacaze and Oefelein~\cite{Lacaze2012}, Banuti \textit{et al.}~\cite{banuti2016a} and Lapenna \textit{et al.}~\cite{lapenna2017a} investigated hydrogen-oxygen and methane-oxygen flames and checked if thermodynamic states in these flames fall underneath the critical locus. All groups~\cite{Lacaze2012,banuti2016a,lapenna2017a} came to the finding that the mixture always stays in a single-phase state and never undergoes a phase separation process. For their investigations they used a framework comprising the cubic EoS for describing the real-gas effects ("dense gas"-approach) and an approximation method for the a-posteriori calculation of the mixture critical locus which reads:
\begin{equation}
T_\textrm{c,mix} = \sum_{j=1}^{N_c} \frac{z_j v_\textrm{c,j}}{\sum_{i=1}^{N_c} z_i v_\textrm{c,i}} T_\textrm{c,j}
\label{eq:Tcmix}
\end{equation}
\begin{equation}
p_\textrm{c,mix} = \frac{\sum_{j=1}^{Nc} \left( z_j \frac{p_\textrm{c,j} v_\textrm{c,j}}{T_\textrm{c,j}}\right) T_\textrm{c,mix}}{\sum_{j=1}^{N_c} z_j v_\textrm{c,j}} \; .
\label{eq:pcmix}
\end{equation}
Here, $v_\textrm{c,i}$, $T_\textrm{c,i}$ and $p_\textrm{c,i}$ are the critical volume, temperature and pressure of the $i$-th species, respectively. According to Poling \textit{et al.}~\cite{poling2001} this approach is a rapid estimate (RE) method and more reliable methods, e.g., Heidemann and Khalil~\cite{heidemann1980a} (HK), are available in the literature which apply the correct physics. The approach of Heidemann and Khalil~\cite{heidemann1980a} is based on a Taylor expansion of the Helmholtz free energy $A$ and the critical point of a mixture is characterized by the following objective functions:
\begin{equation}
\mathbf{Q} \Delta \mathbf{n} = 0 \; ,
\end{equation}
\begin{equation}
\Delta \mathbf{n}^T \Delta \mathbf{n} = 0
\end{equation}
and 
\begin{equation}
C = \sum\limits_{i} \sum\limits_{j} \sum\limits_{k} \Delta n_i \Delta n_j \Delta n_k \frac{\partial^3 A}{\partial n_i \partial n_j \partial n_k}\at[\Bigg]{T,v} = 0 \; .
\end{equation}
In these equations, $\mathbf{n}$ is the vector of mole numbers of the appropriate mixture and $\mathbf{Q}$ is a square matrix with elements:
\begin{equation}
Q_{ij} = \frac{\partial^2 A}{\partial n_i \partial n_j} \at[\Bigg]{T,v} \; .
\end{equation}
$C$ and $\mathbf{Q}$ can be directly obtained from the applied cubic EoS. The calculation of the critical locus is therefore consistent with the thermodynamic framework applied in the LES and is no more an estimation approach.\\

\begin{figure}[t]
	\centering
	\tikzsetnextfilename{comparisonREHKmethod}%
	\input{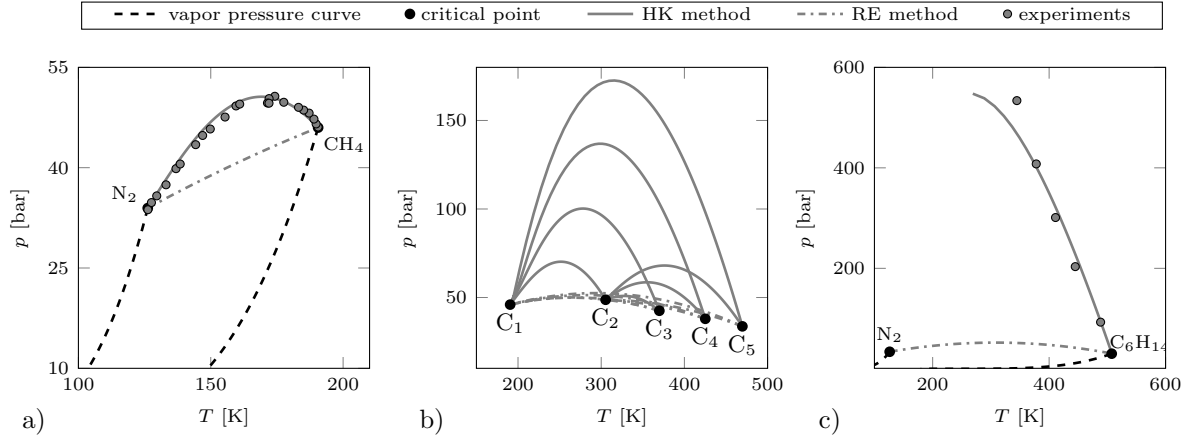}%

	\vspace*{-12pt}
	\caption{Comparison of the rapid estimation (RE) approach and the high fidelity thermodynamic approach by Heidemann and Khalil~\cite{heidemann1980a} (HK) for different binary mixtures: a) nitrogen + methane (experimental data is taken from the literature~\cite{bloomer1953a,cines1953a,chang1967a,stryjek1974a}), b) different alkanes from the homologous series and c) nitrogen + n-hexane (experimental data is taken from the literature~\cite{eliosa2007a}).}
	\vspace*{0pt}
	\label{fig:comparisonREHK}
\end{figure}

\begin{figure}[h]
	\centering
	\tikzsetnextfilename{multicomponentVLE_typeIII}%
	\begin{tikzpicture}
	\begin{axis}
	[
	view={50}{40},	
	width = 0.45\textwidth,
	height = 0.45\textwidth,
	xmin=0,
	xmax=1,
	xlabel = {Mole fraction $x$, $y$, $z$},
	xlabel style = {rotate=-32},
	ymin=70,
	ymax=520,
	ylabel = Temperature $T$,
	ylabel style = {rotate=25},
	zmin=0.2,
	zmax=500,
	zlabel = Pressure $p$,
	x filter/.code={\pgfmathparse{1-#1}\pgfmathresult},
	z filter/.code={\pgfmathparse{#1*1e-5}\pgfmathresult},
	x dir=reverse,
	ticks=none,
	axis lines=left,
	axis line style = ultra thick,
	xlabel style = {font=\small},
	ylabel style = {font=\small},
	zlabel style = {font=\small}
	]	
		\draw (1,70,0.2) -- (1,520,0.2);
		\draw (1,520,0.2) -- (1,520,500);
		\draw (1,70,500) -- (1,520,500);
		\draw (0,520,0.2) -- (1,520,0.2);
		\draw (0,520,500) -- (1,520,500);
		\draw (0,520,0.2) -- (0,520,500);
	
		\addplot3 [color=black, line width=1.5pt,fill=white]  table [x=z, y=T,z=p]{../sources/criticalLocus/Hexane-Nitrogen/VLE_Nitrogen_Hexane_488.4K_kij_0.dat};
		\addplot3 [color=black, line width=1.5pt,fill=white]  table [x=z, y=T,z=p]{../sources/criticalLocus/Hexane-Nitrogen/VLE_Nitrogen_Hexane_411K_kij_0.dat};
		\addplot3 [color=black, line width=1.5pt,fill=white]  table [x=z, y=T,z=p]{../sources/criticalLocus/Hexane-Nitrogen/VLE_Nitrogen_Hexane_310.93K_kij_0.dat};

		\addplot3 [color=black,dashed, line width=2.0pt]  table [x=z, y=T, z expr=\thisrow{p}*1e5]{../sources/criticalLocus/Hexane-Nitrogen/pT_Diagram_Hexane_PR.dat};
		\addplot3 [color=black,dashed, line width=2.0pt]  table [x=z, y=T, z expr=\thisrow{p}*1e5]{../sources/criticalLocus/Hexane-Nitrogen/pT_Diagram_Nitrogen_PR.dat};
		
		\addplot3 [color=black,dashdotted, line width=2.0pt]  table [x=z, y=TScale, z expr=\thisrow{pScale}*1e5]{../sources/criticalLocus/Hexane-Nitrogen/LLG-Line.dat};
			
		\addplot3 [color=gray,line width=2.0pt] table {
		1	507.820000000000	3034000.00000000
		0.990000000000000	507.469699270540	3140056.38079744
		0.980000000000000	507.112079388289	3248319.87315724
		0.970000000000000	506.746847233016	3358862.40734423
		0.960000000000000	506.373757572901	3471756.29628996
		0.950000000000000	505.992554548137	3587076.92545443
		0.940000000000000	505.602971090035	3704902.91661079
		0.930000000000000	505.204728301680	3825316.30213595
		0.920000000000000	504.797534797164	3948402.71059495
		0.910000000000000	504.381085996110	4074251.56447362
		0.900000000000000	503.955063369950	4202956.29098893
		0.890000000000000	503.519133636055	4334614.54698679
		0.880000000000000	503.072947895477	4469328.45902651
		0.870000000000000	502.616140709660	4607204.87984904
		0.860000000000000	502.148329111021	4748355.66253367
		0.850000000000000	501.669111541848	4892897.95376628
		0.840000000000000	501.178066715363	5040954.50777304
		0.830000000000000	500.674752392264	5192654.02261714
		0.820000000000000	500.158704065336	5348131.50071510
		0.810000000000000	499.629433543991	5507528.63560461
		0.800000000000000	499.086427429793	5670994.22718986
		0.790000000000000	498.529145473053	5838684.62790455
		0.780000000000000	497.957018799574	6010764.22247099
		0.770000000000000	497.369447995459	6187405.94419609
		0.760000000000000	496.765801036600	6368791.83103810
		0.750000000000000	496.145411047994	6555113.62500125
		0.740000000000000	495.507573876415	6746573.41877622
		0.730000000000000	494.851545458120	6943384.35394475
		0.720000000000000	494.176538961194	7145771.37551212
		0.710000000000000	493.481721679787	7353972.04802788
		0.700000000000000	492.766211654867	7568237.43910822
		0.690000000000000	492.029073993095	7788833.07678990
		0.680000000000000	491.269316852046	8016039.98783420
		0.670000000000000	490.485887056114	8250155.82486745
		0.660000000000000	489.677665303078	8491496.09110329
		0.650000000000000	488.843460916259	8740395.47235027
		0.640000000000000	487.982006091490	8997209.28708056
		0.630000000000000	487.091949581592	9262315.06653368
		0.620000000000000	486.171849753499	9536114.27816777
		0.610000000000000	485.220166944604	9819034.20726785
		0.600000000000000	484.235255034930	10111530.0131912
		0.590000000000000	483.215352140312	10414086.9785945
		0.580000000000000	482.158570318551	10727222.9720668
		0.570000000000000	481.062884165161	11051491.1469017
		0.560000000000000	479.926118157591	11387482.9013122
		0.550000000000000	478.745932586073	11735831.1282255
		0.540000000000000	477.519807885183	12097213.7859271
		0.530000000000000	476.245027151983	12472357.8242458
		0.520000000000000	474.918656603634	12862043.5046967
		0.510000000000000	473.537523688569	13267109.1570073
		0.500000000000000	472.098192519713	13688456.4187024
		0.490000000000000	470.596936244380	14127056.0088479
		0.480000000000000	469.029705901760	14583954.0915148
		0.470000000000000	467.392095243364	15060279.2888344
		0.460000000000000	465.679300901963	15557250.4073361
		0.450000000000000	463.886077187462	16076184.9441320
		0.440000000000000	462.006684660223	16618508.4407029
		0.430000000000000	460.034831479093	17185764.7505015
		0.420000000000000	457.963606337460	17779627.2807559
		0.410000000000000	455.785401579368	18401911.2564671
		0.400000000000000	453.491824821113	19054587.0323136
		0.390000000000000	451.073597081945	19739794.4411797
		0.380000000000000	448.520435038790	20459858.1092299
		0.370000000000000	445.820914550305	21217303.5765256
		0.360000000000000	442.962312028517	22014873.9239472
		0.350000000000000	439.930419553197	22855546.3993162
		0.340000000000000	436.709328804555	23742548.2250800
		0.330000000000000	433.281177913730	24679370.3076762
		0.320000000000000	429.625854181693	25669776.8814916
		0.310000000000000	425.720644291374	26717808.0981929
		0.300000000000000	421.539822157175	27827771.0491376
		0.290000000000000	417.054162997570	29004212.4325913
		0.280000000000000	412.230370763038	30251862.6643193
		0.270000000000000	407.030405087798	31575536.0914870
		0.260000000000000	401.410694220562	32979964.2033987
		0.250000000000000	395.321223386048	34469526.9563937
		0.240000000000000	388.704496482980	36047829.4421730
		0.230000000000000	381.494388053779	37717043.9855553
		0.220000000000000	373.614941518721	39476896.8243426
		0.210000000000000	364.979246040913	41323117.7422747
		0.200000000000000	355.488669177091	43245083.7119070
		0.190000000000000	345.032989847055	45222270.7018533
		0.180000000000000	333.492454367086	47218995.5741679
		0.170000000000000	320.743590096388	49176851.4433238
		0.160000000000000	306.671855022232	51004427.6384385
		0.150000000000000	291.195686323319	52564885.4903986
		0.140000000000000	274.306968138005	53664742.0729480
		};
		\addplot3 [color=gray,line width=2.0pt,dotted] table {
		0.140000000000000	274.306968138005	53664742.0729480
		0.130000000000000	256.128505194475	54052840.0289673
		0.120000000000000	236.972955171482	53445178.9837945
		0.110000000000000	217.359035464022	51588428.4368520
		0.100000000000000	197.928769040725	48345820.1644726
		};
		\addplot3 [color=gray,line width=2.0pt,smooth] table {
		0		126.192		3395800
		0.025	130.1		4000000
		0.050	133.1		4400000
		0.075	136.4		4200000
		0.1		138.8112	4074960	
		};
		\addplot3 [color=black,mark=*,only marks,line width=1.0pt,mark size = 1.5pt] table {
		0	126.192		3395800
		0.1	138.8112	4074960	
		1	507.820		3034000
		};
		
		\draw (axis cs:0.0,507.820,30.34) -- (axis cs:0.15,510,120) node [anchor=south] {CP$_\text{HV}$};
		\draw (axis cs:1.0,126.192,33.95) -- (axis cs:1.0,140,60) node [anchor=south] {CP$_\text{LV}$};
		\draw (axis cs:0.95,133,44) -- (axis cs:0.87,173,44) node [anchor=west] {LG-Eq.};
		\draw (axis cs:0.89,100,14) -- (axis cs:0.8,115,18) node [anchor=west] {LLG-Eq.};
		\draw (axis cs:0.79,364.9,413.23) -- (axis cs:0.70,450,413.23) node [anchor=west] {LG-Eq.};
	\end{axis}		
\end{tikzpicture}%

	\caption{Principal sketch of a binary mixture with a type III critical locus.}
	\label{fig:multicomponentVLE_typeIII}
\end{figure}
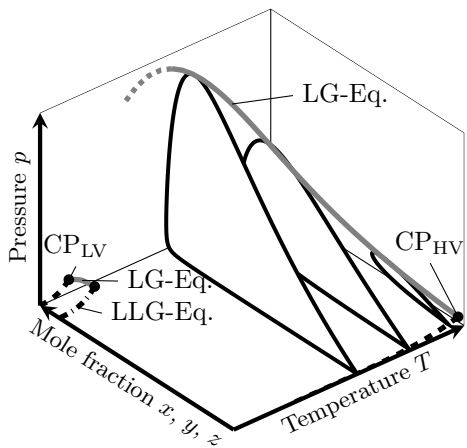

In Fig.~\ref{fig:comparisonREHK} the RE method is compared to the HK method for different (non-reacting) binary mixtures. A binary mixture of nitrogen (N$_2$) and methane (CH$_4$) is shown in Fig.~\ref{fig:comparisonREHK}~a which forms a critical locus of type I similar to the sketch in Fig.~\ref{fig:multicomponentVLE_typeI}. A clear difference can be seen between the two methods in terms of pressure values of the critical locus. As the high fidelity method of Heidemann and Khalil~\cite{heidemann1980a} has a very good agreement with experimental data it can be concluded that the RE method underpredicts the critical locus. This fact can be substantiated by comparing more different types of molecules. In Fig.~\ref{fig:comparisonREHK}~b the critical loci of the binary mixtures of various alkanes of the homologous series (methane C$_1$ to n-pentane C$_5$) are shown. A significant underestimation of the RE method becomes obvious up to a factor of more than three for the binary mixture of methane and n-pentane. The RE method shows almost no superelevation with respect to the pressure values of the critical locus and hence the component with the highest critical pressure determines the maximum of the critical locus. This fact gets even worse if mixtures are investigated which do not form a type I critical locus. In Fig.~\ref{fig:comparisonREHK}~c a binary mixture of nitrogen (N$_2$) and n-hexane (C$_6$H$_{14}$) is plotted. This mixture has a critical locus of type III~\cite{vanKonynenburg1980a} which is characterized by very large pressure values for the liquid gas equilibrium and a not complete miscibility of the two components~\cite{kay1968a}. Therefore, in type III mixtures a second liquid phase can appear and lead to a liquid-liquid-gas (LLG) equilibrium which is illustrated in Fig.~\ref{fig:multicomponentVLE_typeIII} by a principal sketch. In addition, the extension of the state space with phase separation phenomena is drastically enlarged compared to the type I mixture, compare Fig.~\ref{fig:multicomponentVLE_typeI}. In Fig.~\ref{fig:comparisonREHK}~c, the LG branch of the critical locus emerging from the high volatile component (C$_6$H$_{14}$) is plotted and compared to experimental data from the literature~\cite{eliosa2007a}. The high fidelity approach gives an excellent prediction up to a pressure of approximately 400~bar. The RE method completely fails in predicting the critical locus. We therefore conclude that it is highly important to use a thermodynamically consistent approach rather than a RE method to predict phase equilibria in multicomponent mixtures. Similar findings have been reported by Banuti \textit{et al.}~\cite{banuti2017a}.

\subsubsection{Thermodynamically Consistent Approach}

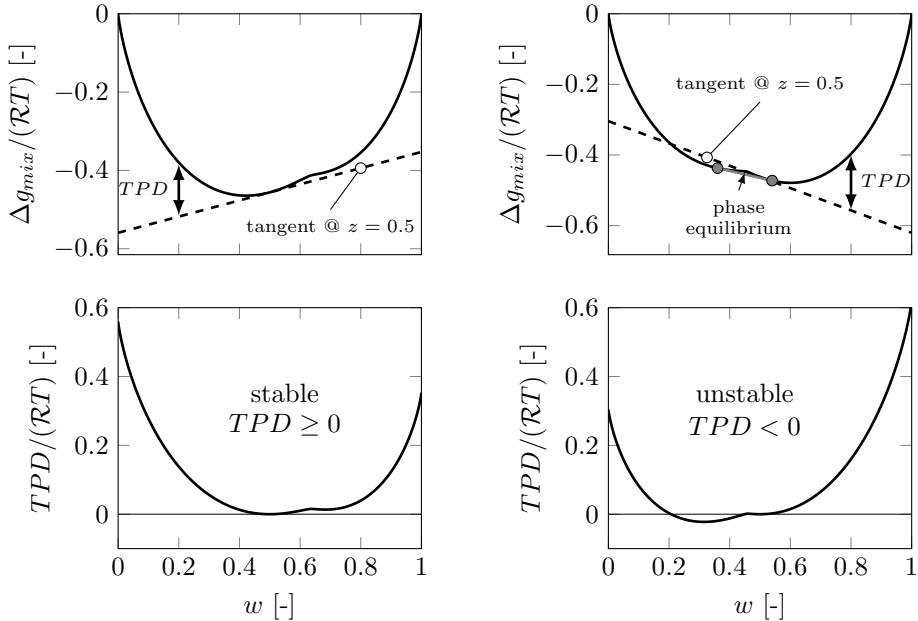
\begin{figure}[h]
	\centering
	\tikzsetnextfilename{TPD}%

\begin{tikzpicture}
	\begin{groupplot}[group style = {group name = plots,group size = 2 by 2, horizontal sep = 70pt, vertical sep = 20pt}, width = 0.35\textwidth, height = 0.3\textwidth]
		\nextgroupplot
		[
		xmin=0,
		xmax=1,
		xticklabels={,,}
		ymin = -0.8,
		ymax = 0,
		ylabel = $\Delta g_{mix} / (\mathcal{R} T)\text{ [-]}$,
		legend style={legend pos=south east,font=\tiny,draw=none}
		]
			\addplot [solid,line width=1.0pt,color=black] table [x=W,y=deltag]{../sources/TPD/TPD_Methane_Nitrogen_40bar_150K_z0.5.dat};
			\addplot [dashed,line width=1.0pt,color=black] table [x=W,y=tangent]{../sources/TPD/TPD_Methane_Nitrogen_40bar_150K_z0.5.dat};
			\draw[latex-latex,line width=1.0pt,color=black] (axis cs: 0.2,-0.385) -- (axis cs: 0.2,-0.515) node [anchor = south east,yshift=0.9ex] {\scriptsize $TPD$};
			\draw [-] (axis cs:0.8,-0.3941)--(axis cs:0.7,-0.5) node[anchor=north,font=\scriptsize,align=center] {tangent @ $z=0.5$} ;
			\draw [only marks, mark=*,fill=white] plot coordinates 
			{
				(axis cs:0.8,-0.3941)
			};
		\nextgroupplot
		[
		xmin=0,
		xmax=1,
		xticklabels={,,}
		ymin = -0.8,
		ymax = 0,
		ylabel = $\Delta g_{mix} / (\mathcal{R} T)\text{ [-]}$,
		legend style={legend pos=south east,font=\tiny,draw=none}
		]
			\addplot [solid,line width=1.0pt,color=black] table [x=W,y=deltag]{../sources/TPD/TPD_Methane_Nitrogen_40bar_160K_z0.5.dat};
			\addplot [dashed,line width=1.0pt,color=black] table [x=W,y=tangent]{../sources/TPD/TPD_Methane_Nitrogen_40bar_160K_z0.5.dat};
			\draw[latex-latex,line width=1.0pt] (axis cs: 0.8,-0.400) -- (axis cs: 0.8,-0.555) node [anchor = south west,yshift=1.1ex] {\scriptsize $TPD$};
			\draw [-] (axis cs:0.325,-0.4069)--(axis cs:0.5,-0.25) node[anchor=south,font=\scriptsize,align=center] {tangent @ $z=0.5$} ;
			\draw [only marks, mark=*,fill=white] plot coordinates 
			{
				(axis cs:0.325,-0.4069)
			};
			\draw [line width=1.0pt,color=gray] (axis cs:0.360,-0.4372)--(axis cs:0.540,-0.4726);
			\draw [only marks, mark=*,fill=gray] plot coordinates 
			{
				(axis cs:0.360,-0.4372)
				(axis cs:0.540,-0.4726)
			};
			\draw [latex-] (axis cs:0.45,-0.45) -- (axis cs:0.43,-0.50) node [anchor=north,align=center,font=\scriptsize] {phase \\ equilibrium};
		\nextgroupplot
		[
		xmin=0,
		xmax=1,
		xlabel = $w \text{ [-]}$,
		ymin=-0.1,
		ymax = 0.6,
		ylabel = $TPD / (\mathcal{R} T)\text{ [-]}$,
		legend style={legend pos=south east,font=\tiny,draw=none}
		]
			\addplot [line width=0.1pt] table 
			{
				0.0	0.0
				1.0	0.0
			};
			\addplot [solid,line width=1.0pt,color=black] table [x=W,y=TPD]{../sources/TPD/TPD_Methane_Nitrogen_40bar_150K_z0.5.dat};

			\node at (axis cs:0.55,0.3) [anchor=south] {stable};
			\node at (axis cs:0.55,0.2) [anchor=south] {$TPD \ge 0$};
		\nextgroupplot
		[
		xmin=0,
		xmax=1,
		xlabel = $w \text{ [-]}$,
		ymin=-0.1,
		ymax = 0.6,
		ylabel = $TPD / (\mathcal{R} T)\text{ [-]}$,
		legend style={legend pos=south east,font=\tiny,draw=none}
		]
			\addplot [line width=0.1pt] table 
			{
				0.0	0.0
				1.0	0.0
			};
			\addplot [solid,line width=1.0pt,color=black] table [x=W,y=TPD]{../sources/TPD/TPD_Methane_Nitrogen_40bar_160K_z0.5.dat};

			\node at (axis cs:0.45,0.3) [anchor=south] {unstable};
			\node at (axis cs:0.45,0.2) [anchor=south] {$TPD < 0$};
	\end{groupplot}	
\end{tikzpicture}%

	\caption{Visualization of the tangent plane distance (TPD) method for the binary mixture methane-nitrogen at a pressure of 40~bar and two different temperatures (left: 150~K, right: 160~K). As feed composition a equimolar mixture ($z = 0.5$) was taken. }
	\label{fig:TPD}
\end{figure}

In order to take multicomponent phase separation into account during the simulation, it would be possible to tabulate the VLE region a-priori. This, however, is not really practical especially when the model is applied to mixtures that have more than two components which is the case in combustion. Hence, we recently implemented and validated a vapor liquid equilibrium (VLE) model, see Traxinger \textit{et al.}~\cite{traxinger2017a,traxinger2018a}, which checks the stability of the single phase mixture and performs a phase separation in the case of an unstable mixture. The model was inspired by the work of Qiu and Reitz~\cite{qiu2015a} and was also applied successfully by an other group~\cite{matheis2018a}. To check the stability of the single-phase mixture the tangent plane distance (TPD) method of Michelsen~\cite{michelsen1982a} is used which reads:
\begin{equation}
TPD \left( \mathbf{w} \right) = \sum_{i} w_i \left[ \ln w_i  + \ln \varphi_i \left(\mathbf{w}\right) - \ln z_i  - \ln \varphi_i \left(\mathbf{z}\right) \right] \; .
\label{eq:TPD}
\end{equation}
Here, $\mathbf{w} = \{ w_1 , ... , w_{N_c} \}$ is a trial phase composition and $\varphi_i$ is the fugacity coefficient of component $i$ which can be directly obtained from any cubic equation of state, for more details see, e.g., Michelsen~and~Mollerup~\cite{michelsen2007} or Firoozabadi~\cite{firoozabadi1999}. In the case of $TPD$ being smaller than zero for any $\mathbf{w}$, the mixture is considered unstable, a phase separation is performed and a flash is solved yielding the appropriate phase compositions. The concept of the TPD method is visualized in Fig.~\ref{fig:TPD}, where a stable (left) and an unstable (right) mixture is shown. In the unstable case the tangent aligned to the Gibbs-Energy surface (here: the surface reduces to a single curve) at the corresponding feed composition intersects the curve of the Gibbs-Energy. The corresponding phase equilibrium at the appropriate pressure and temperature is illustrated by the gray dots and the gray line in Fig.~\ref{fig:TPD}~right. In the present study, we consider a maximum of two phases. An instantaneous separation process is assumed and therefore possible non-equilibrium effects like they occur for instance in steam turbines~\cite{gyarmathy1962a} are neglected. Furthermore, only multicomponent phase separation is investigated as the injection pressure is usually well above the critical pressure of the pure components (fuel and oxidizer) and we do not expect any strong expansion effects in the region where only a pure fluid is present.

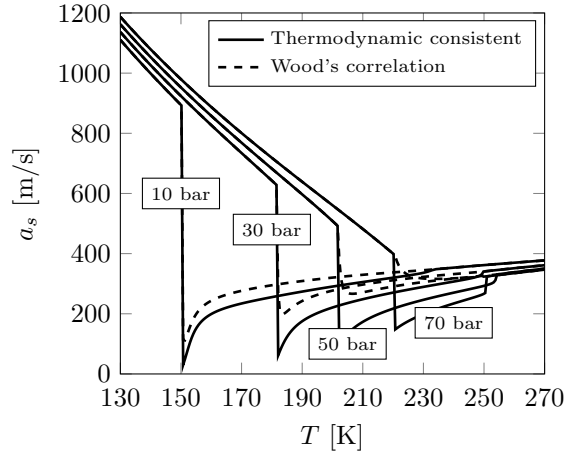
\begin{figure}[h]
	\centering
	\tikzsetnextfilename{prudhoeBayAs}%

\begin{tikzpicture}
	\begin{axis}
	[
	width = 0.45\textwidth, 
	height = 0.40\textwidth,
    xmin=130,
    xmax=270,
    xtick = {130,150,170,190,210,230,250,270},
    xlabel = $T\text{ [K]}$,
    x tick label style={yshift=-2pt},
    ymin=0,
    ymax=1200,
    ylabel = $a_s\text{ [m/s]}$,
    ytick = {0,200,400,600,800,1000,1200,1400},
    /pgf/number format/1000 sep={},
    legend style={font=\scriptsize,anchor=north east,legend cell align=left}
    ]
	    \addplot [color=black,line width=1.0pt] table [x=T, y=as]{../sources/prudhoeBay/as_PrudhoeBay_p_10bar.dat};\addlegendentry{Thermodynamic consistent};
	    \addplot [color=black,line width=1.0pt,dashed] table [x=T, y=asWood]{../sources/prudhoeBay/as_PrudhoeBay_p_10bar.dat};\addlegendentry{Wood's correlation};
	    \node [draw=black,fill=white,font=\scriptsize] at (axis cs: 150,600) {10~bar};
 	    \addplot [color=black,line width=1.0pt,forget plot] table [x=T, y=as]{../sources/prudhoeBay/as_PrudhoeBay_p_30bar.dat};
 	    \addplot [color=black,line width=1.0pt,forget plot,dashed] table [x=T, y=asWood]{../sources/prudhoeBay/as_PrudhoeBay_p_30bar.dat};
 	    \node [draw=black,fill=white,font=\scriptsize] at (axis cs: 180,480) {30~bar};
   	    \addplot [color=black,line width=1.0pt,forget plot] table [x=T, y=as]{../sources/prudhoeBay/as_PrudhoeBay_p_50bar.dat};
   	    \addplot [color=black,line width=1.0pt,forget plot,dashed] table [x=T, y=asWood]{../sources/prudhoeBay/as_PrudhoeBay_p_50bar.dat};
   	    \node [draw=black,fill=white,font=\scriptsize] at (axis cs: 205,100) {50~bar};
   	    \addplot [color=black,line width=1.0pt,forget plot] table [x=T, y=as]{../sources/prudhoeBay/as_PrudhoeBay_p_70bar.dat};
   	    \addplot [color=black,line width=1.0pt,forget plot,dashed] table [x=T, y=asWood]{../sources/prudhoeBay/as_PrudhoeBay_p_70bar.dat};
   	    \node [draw=black,fill=white,font=\scriptsize] at (axis cs: 240,170) {70~bar};
	\end{axis}	
\end{tikzpicture}%

	\caption{Comparison of the consistently calculated speed of sound $a_s$ and the speed of sound correlation of Wood~\cite{wood1930a} at four different pressure levels (compare Fig.~\ref{fig:prudhoeBayVLE}) for the 14 component Prudhoe Bay mixture\cite{picard1987a}. The Peng-Robinson EoS~\cite{peng1976a} was applied for this calculation.}
	\label{fig:prudhoeBayAs}
\end{figure}

Similar to pure fluids, it is important to keep in mind that the fluid properties can be subject to large changes after the VLE is entered. For mass/mole specific properties the respective vapor and liquid properties are weighted with the appropriate vapor fraction $\beta$ to arrive at the representative two-phase value. In terms of the density this reads:
\begin{equation}
\rho = \left( 1 - \beta \right) \rho^\textrm{l} + \beta \rho^\textrm{v} \; .
\label{eq:rhoTwoPhase}
\end{equation}
As a result of this weighting approach, the change of these kind of properties is continuous throughout the VLE. For the sake of simplicity, we will apply the same approach in this study for the transport properties~\cite{awad2008a}. For derived properties like the compressibility $\psi$ and the speed of sound $a_s$ it is important to use a fully consistent calculation to avoid solver instabilities. Weighting the vapor and liquid properties can lead to severe deviations from the consistently calculated properties~\cite{nichita2010a}. A commonly applied weighting approach for the calculation of the speed of sound is the correlation proposed by Wood~\cite{wood1930a}. This correlation is able to reproduce the basic tendencies inside the two-phase region, but is not suitable to predict the sudden drop occurring after the crossing of the dew- and bubble-point line and hence the correct absolute values, see Fig.~\ref{fig:prudhoeBayAs}. We therefore recently extended our thermodynamic framework such, that it is able to consistently predict the speed of sound and the compressibility inside the VLE. For more details, the complete derivation as well as a thorough validation the interested reader is referred to Traxinger \textit{et~al.}~\cite{traxinger2018a}. 

\subsubsection{Validation with Experimental Data}

\begin{figure}[h]
	\centering
	\tikzsetnextfilename{validationVLE}%
	\input{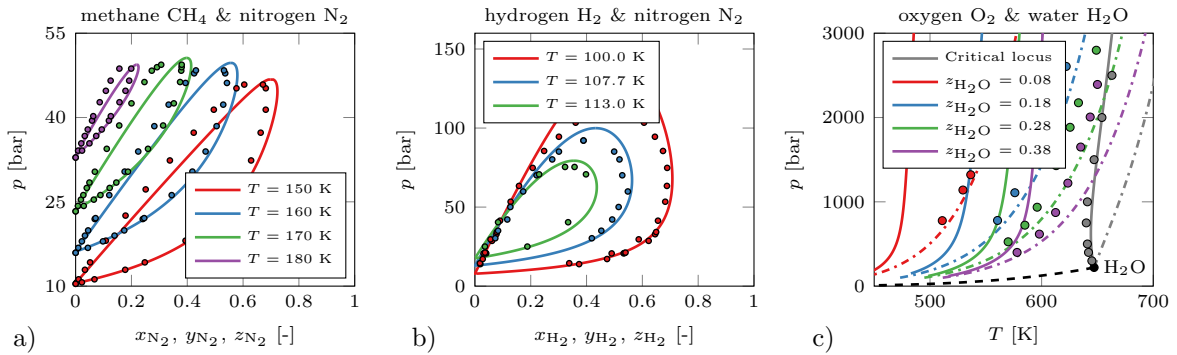}%

	\caption{Validation of the vapor-liquid equilibrium calculations with experimental data~\cite{kidnay1975a,shtekkel1939a,streett1978a,japas1985a} for three different binary mixtures: a) methane and nitrogen ($k_{ij} = 0$), b) hydrogen and nitrogen ($k_{ij} = 0$) and oxygen and water (solid: $k_{ij} = 0$; dash-dotted: $k_{ij} = 0.3$).}
	\label{fig:validationVLE}
\end{figure}

In Fig.~\ref{fig:comparisonREHK} we already have shown some first comparison between predicted and experimentally determined critical loci for type I and type III mixtures. The overall conclusion was that the SRK-EoS in combination with the high fidelity method of Heidemann and Khalil~\cite{heidemann1980a} is able to give good predictions of the critical locus of different types of mixtures over a wide pressure range. To check the prediction quality of the SRK-EoS in the mixture space and at more relevant conditions, additional experimental data available in the literature has been used. In the Figs.~\ref{fig:validationVLE}~a and b isothermal VLEs for the binary mixture of the two different fuels (methane and hydrogen) with nitrogen are shown. For both mixtures, the binary interaction parameter $k_{ij}$ has been set to zero and the SRK-EoS gives good results in the relevant pressure range ($p$~$<$~80~bar). The third binary mixture we are using for our validation is oxygen - water. This mixture occurs in the core of the flame where the cryogenic oxygen mixes with the combustion products. Water is a main combustion product and has a very high critical point ($T_c$~=~647.1~K and $p_c$~=~217.7~bar) as well as strong hydrogen bondings.
Due the strongly associating character of water cubic EoS are usually not the optimal choice for a good prediction~\cite{aparicio2007a}. However, the cubic EoS are state of the art in the rocket community due to their relative good accuracy for many fluids and mixtures and of course due to their efficiency. In Fig.~\ref{fig:validationVLE}~c predicted and experimental data are compared for four different isopleths and the critical locus up to 3000~bar. The solid lines correspond to $k_{ij} = 0$ and the dash-dotted lines to $k_{ij} = 0.3$. The water-oxygen mixture forms a critical locus of type III. The comparison of the different binary interaction coefficients shows that a perfect fit of the experimental data by changing $k_{ij}$ is not feasible. Using $k_{ij} = 0$ the critical locus can be predicted very good. By applying $k_{ij} = 0.3$ the prediction of the isopleths gets far better but the calculated critical locus shows large deviations from the experimental data. Furthermore, it has to be noticed that all the experimental data available in the literature~\cite{japas1985a} is for very large pressures ($p > 220$~bar). The operating conditions for our present investigations are however below 80~bar. We therefore decided to perform a sensitivity analyses for the binary interaction coefficient of water-oxygen within the flamelet calculation to see its influence on the result. Four different binary interaction coefficients ($k_{ij} = \left[ 0.0, 0.1, 0.2, 0.3\right]$) have been used for this analysis and no significant change in the results has been found. Therefore, we decided to use $k_{ij} = 0.3$ as binary interaction coefficient between water and oxygen as it fits the curve for $z_{\textrm{H}_2\textrm{O}} = 0.08$ in Fig.~\ref{fig:validationVLE}~c very good and phase separation can only be expected at moderate temperature and hence low water mole fractions. 

\subsection{Real-Gas Flamelet Model}
Assuming that the chemical time scales are smaller than the turbulent time scales, a turbulent non-premixed flame can be represented by a brush of laminar counterflow diffusion flames. In an axisymmetric configuration, the governing equations reduce to an one-dimensional problem that can be solved in mixture fraction space $f$ by employing a coordinate transformation. Under the assumption of a unitary Lewis-number for all species (Le$_i$=Le=1), the so-called flamelet equations read~\cite{Peters2000,Kim2013}:
\begin{gather} \label{eq:flamelet equations}
	\rho \frac{\partial  Y_i}{\partial t} = \rho \frac{\chi}{2} \frac{\partial^2 Y_i}{\partial f^2} + \dot\omega_i \; , \\
	\rho \frac{\partial h}{\partial t} = \rho \frac{\chi}{2} \frac{\partial^2 h}{\partial f^2} \; .
\end{gather}
Here, $\chi$ denotes the scalar dissipation rate, which acts as an inverse diffusion time being an external parameter. $Y_i$ and $\dot\omega_i$ are the mass fraction and chemical source term of species $i$, respectively. The equations are closed using the thermodynamics model presented above and a suitable reaction mechanism for the respective propellant combination, see Sec.~\ref{sec: Theoretical Background and Models}.\ref{sec: Chemical Kinetics}. The flamelet equations are solved for a range of scalar dissipation rates up to extinction. The gas composition is assumed to be independent of pressure, while for all other thermodynamic properties pressure effects are considered~\cite{Zips2018}. Favre presumed probability density functions (PDF) are employed to calculate the filtered values of the laminar database. Under the assumption of statistical independence of the control variables, a $\beta$-PDF is used to approximate the PDF of the mixture fraction and Dirac functions are employed for the PDFs of scalar dissipation rate and pressure. Consequently, the thermo-chemical library is characterized by four control parameters ($f$, $f''^2$, $\chi$, $p$), which need to be provided by the LES.

In order to retrieve the local species composition in the LES, an additional transport equation is solved for the filtered mixture fraction
\begin{equation}
\frac{\partial}{\partial t} \left(  \bar{\rho} \tilde{f} \right)+ \frac{\partial}{\partial x_i} \left(\bar{\rho} \tilde{u}_i \tilde{f} \right) = \frac{\partial}{\partial x_i} \left( \left( \frac{\bar{\mu}}{Sc} + \frac{\mu_{sgs}}{Sc_t}\right)\frac{\partial \tilde{f}}{\partial x_i}\right).
\label{eq:mixture fraction}
\end{equation}
A transport equation for its variance is solved to model the unresolved fluctuations of the mixture fraction, cf. Kemenov \textit{et al.}~\cite{Kemenov2012}:
\begin{equation}
\begin{split}
\frac{\partial}{\partial t} \left(\partial \bar{\rho} \widetilde{f''^2}\right)+ 
\frac{\partial}{\partial x_i} \left( \bar{\rho} \tilde{u}_i \widetilde{f''^2} \right)= \\
\frac{\partial}{\partial x_i} \left( \left( \frac{\bar{\mu}}{Sc} + \frac{\mu_{sgs}}{Sc_t}\right)\frac{\partial \widetilde{f''^2}}{\partial x_i}\right)- 
2 \bar{\rho} \tilde{\chi} + 2 \left(\frac{\bar{\mu}}{Sc} + \frac{\mu_{sgs}}{Sc_{t}}\right) \left( \frac{\partial \tilde{f}}{\partial x_i}\right)^2.
\label{eq:mixture fraction variance}
\end{split}
\end{equation}
The SGS viscosity $\mu_{sgs}$ is evaluated using the model of Vreman~\cite{Vreman2004} with the original constant $c_v=0.07$. The scalar dissipation rate $\chi$ is decomposed into resolved and SGS contributions~\cite{Domingo2008} as
\begin{equation}
2 \bar{\rho} \tilde{\chi} = 2 \frac{\tilde{\mu}}{Sc} \left( \frac{\partial \tilde{f}}{\partial x_i}\right)^2 + C_{\chi} \frac{\mu_{sgs}}{Sc_t} \frac{\widetilde{f''^2}}{\Delta^2},
\label{eq:chi}
\end{equation}
where $\Delta$ is the local filter size and the model constant is set to $C_{\chi}=2$ according to Kemenov \textit{et al.}~\cite{Kemenov2012}. In Eqs.~\eqref{eq:mixture fraction} and \eqref{eq:mixture fraction variance}, the gradient diffusion hypothesis is used to describe the turbulent scalar flux. In particular, the molecular ($Sc$) and turbulent ($Sc_t$) Schmidt numbers are introduced to connect the diffusion coefficients to the viscosity with the required constants being $Sc = Sc_t = 0.7$ in all simulations. Localized artificial dissipation $\mu^*$ based on Cook and Cabot~\cite{Cook2005} 
\begin{equation}
 \mu^* = C_{ad} \: \bar{\rho} \, a_s \, \Delta^2 \, \left|\frac{\partial Z}{\partial x_i}\right|,
\end{equation}
is added to the effective viscosity in Eqs.~\eqref{eq:mixture fraction} and \eqref{eq:mixture fraction variance} in regions of high density gradients to avoid spurious oscillations as $\mu = \mu_m + \mu^*$. Here, $\mu_m$ is the molecular viscosity evaluated according to the empirical model of Chung \textit{et al.}~\cite{Chung1988}. $C_{ad}$ is a user-specified constant which has been set to 0.04, $a_s$ is the speed of sound and $\Delta$ denotes the local filter width. The sensor, namely the local gradient of the compressibility factor ($Z = p v / \mathcal{R} T$), ensures that the effect of the artificial dissipation is limited to the narrow region of strong density gradients due to real-gas thermodynamics.

\subsection{Chemical Kinetics} \label{sec: Chemical Kinetics}
The solution of the flamelet equations requires a chemical reaction mechanism to determine the species source terms. Here, the mechanism of O'Connaire \textit{et al.}~\cite{Connaire2004} is used for the LOx/H$_2$ flame. The mechanism contains nine species and 19 reactions and has been used for the analysis of high-pressure hydrogen flames by Lacaze and Oefelein~\cite{Lacaze2012}. In the case of LOx/CH$_4$ combustion the GRI-3.0~\cite{book:GRI30} mechanism involving 53 species and 325 reactions is employed, which has also been used by Kim \textit{et al.}~\cite{Kim2013} for similar operating conditions.

\subsection{Solver}
The open-source software toolbox OpenFOAM has been extended by the required combustion and thermodynamic models and is employed for all simulations. The unstructured finite-volume code is pressure-based and uses a PISO algorithm to solve the governing equations. In order to consistently incorporate real-gas thermodynamics, the algorithm is modified according to Jarczyk and Pfitzner~\cite{Jarczyk2012}. Second-order accurate central differences are used for spatial discretization and time integration is performed using a first-order implicit Euler scheme. Within the LES, a TVD limiter is additionally employed to the convective term. The basic computational framework has been previously used to simulate flows under rocket-like and engine-relevant conditions.~\cite{muller2016a,traxinger2018a,Zips2018} 


\section{Results and Discussion}\label{sec:Results}

The present study consists of two steps: First, one-dimensional counterflow diffusion flames of both propellant combinations are investigated in Sec.~\ref{sec:Results}.\ref{subsec:Flame Structure Analysis}. Second, the results of the methane flame are tabulated and employed to perform a LES of a well-known reference experiment described in Sec.~\ref{sec:Results}.\ref{subsec: Test Cases and Setup}. The presentation and discussion of the LES results will be carried out in Sec.~\ref{sec:Results}.\ref{subsec: Results}.

\subsection{Counterflow Diffusion Flames} \label{subsec:Flame Structure Analysis}

\subsubsection{General Discussion}

The reference conditions for the counterflow diffusion flames are taken from two well-known experiments. One is the experiment of Singla \textit{et al.}~\cite{Singla2005} at the Mascotte test bench where methane is used as fuel. The other one is a model combustor named BKH~\cite{Hardi2014b,Hardi2014c} operated at DLR Lampoldshausen which is fired with hydrogen. Both combustors are operated under rocket-relevant conditions. This implies that the pressure is supercritical with respect to the propellants' critical point and that the oxidizer is injected at a liquid-like (high density) state, hence cryogenic temperatures. An overview of the actual operating conditions is given in Tab.~\ref{tab:operating conditions}. For both nominal operating conditions laminar flamelet calculations have been performed with a scalar dissipation rate $\chi=1$.

\begin{table}[h]
\centering
\caption{\label{tab:operating conditions} Operating conditions of the considered LOx/H$_{2}$ and LOx/CH$_{4}$ flames.}
\begin{tabular}{cccccccc}
	\multicolumn{8}{c}{LOx/H$_{2}$ flame} 	\\ \hline 
	$p_{op}$ [MPa] & $T_{\textrm{O}_2}$ [K] & $T_{\textrm{H}_2}$ [K] & $U_{\textrm{O}_2}$ [m/s] & $U_{\textrm{H}_2}$ [m/s] & $\rho_{\textrm{O}_2}$  [kg/m$^3$] & $\rho_{\textrm{H}_2}$ [kg/m$^3$] &  $J$ [-]  \\ 
	$6.07$	&	$127$	 & $279$ & $12.3$ & $410$ & $926.27$ &  $5.10$ & 6.12 	\\ \hline
	\multicolumn{8}{c}{LOx/CH$_{4}$ flame} \\ \hline
	$p_{op}$ [MPa] & $T_{\textrm{O}_2}$ [K] & $T_{\textrm{CH}_4}$ [K]  & $U_{\textrm{O}_2}$ [m/s]  & $U_{\textrm{CH}_4}$ [m/s] & $\rho_{\textrm{O}_2}$  [kg/m$^3$] & $\rho_{\textrm{CH}_4}$ [kg/m$^3$] &  $J$ [-] \\ 
	$5.61$	&	$85$	 & $288$ & $2.48$ & $63.20$	& $1180.08$ & $41.79$ & 23.00	\\ \hline
\end{tabular}
\end{table}

\begin{figure}[!h]
	\centering
	\tikzsetnextfilename{flameletBase}%

\begin{tikzpicture}
	\begin{groupplot}[group style = {group name = plots,group size = 2 by 1, horizontal sep = 90pt}, width = 0.26\textwidth, height = 0.26\textwidth, scale only axis]
	\nextgroupplot[
		xmajorgrids,
		ymajorgrids,
		xmin=0,
		xmax=1,
		xtick={ 0,0.2,0.4,0.6,0.8,1},
		xlabel={$f$ [-]},
		x tick label style={yshift=-2pt},
		ymin=0,
		ymax=4000,
		ytick={0, 1e3, 2e3, 3e3, 4e3},
		/pgf/number format/1000 sep={},
		ylabel={$T$ [K]},
		clip=false,
		legend style={font=\small,anchor=north east,legend cell align=left}
		]
		
		\addplot [color=black,solid,line width=1pt,smooth] table [y=T, x=x, col sep=comma]{../sources/Flamelets/BKH_Connaire_chi1_kij_0.3/postProcessing/sampleDict_thermo/0.01240101/thermo_T_ha_hs_rho_Cp_thermo:alpha_thermo:mu_HeatRelease_thermo:Z_thermo:psi_thermo:as_betaN.csv};
		\addlegendentry{LOx/H$_2$};
		\addplot [color=black,dashed,line width=1pt,smooth] table [y=T, x=x, col sep=comma]{../sources/Flamelets/Mascotte_GRI_RG_chi1_kij_0.3/postProcessing/sampleDict_thermo/0.00026562/thermo_T_ha_hs_rho_Cp_thermo:alpha_thermo:mu_HeatRelease_thermo:Z_thermo:psi_thermo:as_betaN.csv};
		\addlegendentry{LOx/CH$_4$};
		
		
	\nextgroupplot[
		xmajorgrids,
		ymajorgrids,
		xmode = log,
		xmin=0.0001,
		xmax=1,
		xtick={ 0.0001,  0.001, 0.01, 0.1, 1},
		xlabel={$f$ [-]},
		x tick label style={yshift=-2pt},
		ymin=0,
		ymax=1200,
		ylabel={$\rho$ [kg/m$^3$]},
		/pgf/number format/1000 sep={},
		legend style={font=\scriptsize,anchor=north east,legend cell align=left}
		]
		\addplot [color=black,solid,line width=1pt,smooth] table [y=rho, x=x, col sep=comma]{../sources/Flamelets/BKH_Connaire_chi1_kij_0.3/postProcessing/sampleDict_thermo/0.01240101/thermo_T_ha_hs_rho_Cp_thermo:alpha_thermo:mu_HeatRelease_thermo:Z_thermo:psi_thermo:as_betaN.csv};		
		\addplot [color=black,dashed,line width=1pt,smooth] table [y=rho, x=x, col sep=comma]{../sources/Flamelets/Mascotte_GRI_RG_chi1_kij_0.3/postProcessing/sampleDict_thermo/0.00026562/thermo_T_ha_hs_rho_Cp_thermo:alpha_thermo:mu_HeatRelease_thermo:Z_thermo:psi_thermo:as_betaN.csv};		
	\end{groupplot}
\begin{groupplot}[group style = {group name = plots,group size = 2 by 1, horizontal sep = 90pt}, width = 0.26\textwidth, height = 0.26\textwidth, scale only axis,axis x line=none,ytick pos=right]
	\nextgroupplot[
		xmin=0,
		xmax=1,
		xtick={ 0,0.2,0.4,0.6,0.8,1},
		xlabel={},
		ymin=0,
		ymax=4000,
		yticklabels={,,},
		ylabel={}
		]
	\nextgroupplot[
		xmode = log,
		xmin=0.0001,
		xmax=1,
		xtick={ 0.0001,  0.001, 0.01, 0.1, 1},
		xlabel={$f$ [-]},
		ymin=0,
		ymax=1.2,
		ylabel={$Z$ [-]},
		]
		\addplot [color=black,solid,line width=1pt,smooth] table [y=thermo:Z, x=x, col sep=comma]{../sources/Flamelets/BKH_Connaire_chi1_kij_0.3/postProcessing/sampleDict_thermo/0.01240101/thermo_T_ha_hs_rho_Cp_thermo:alpha_thermo:mu_HeatRelease_thermo:Z_thermo:psi_thermo:as_betaN.csv};
		\addplot [color=black,dashed,line width=1pt,smooth] table [y=thermo:Z, x=x, col sep=comma]{../sources/Flamelets/Mascotte_GRI_RG_chi1_kij_0.3/postProcessing/sampleDict_thermo/0.00026562/thermo_T_ha_hs_rho_Cp_thermo:alpha_thermo:mu_HeatRelease_thermo:Z_thermo:psi_thermo:as_betaN.csv};
		\draw[color=black,line width=0.5pt,rotate around={-15:(axis cs:0.0045,0.95)}] (axis cs:0.0045,0.95) ellipse (0.50cm and 0.23cm);
		\draw[color=black,line width=0.5pt,-latex] (axis cs:0.0125,0.91) -- (axis cs:0.03,0.85) -- (axis cs:0.6,0.85);
		\draw[color=black,line width=0.5pt] (axis cs:0.004,0.08) ellipse (0.60cm and 0.23cm);
		\draw[color=black,line width=0.5pt,-latex] (axis cs:0.0011,0.08) -- (axis cs:2e-4,0.08);
	\end{groupplot}
   	\node[below = 0.4cm of plots c1r1.south,xshift=-3.5cm] {a)};
   	\node[below = 0.4cm of plots c2r1.south,xshift=-3.5cm] {b)};
\end{tikzpicture}%

	\vspace*{-0pt}
	\caption{Results of the flamelet calculations for the LOx/H$_2$ and the LOx/CH$_4$ flames at $\chi=1$ and the conditions according to Tab.~\ref{tab:operating conditions}: a) temperature profiles; b) density and compressibility factor profiles.}
	\label{fig:flameletBase}
\end{figure}
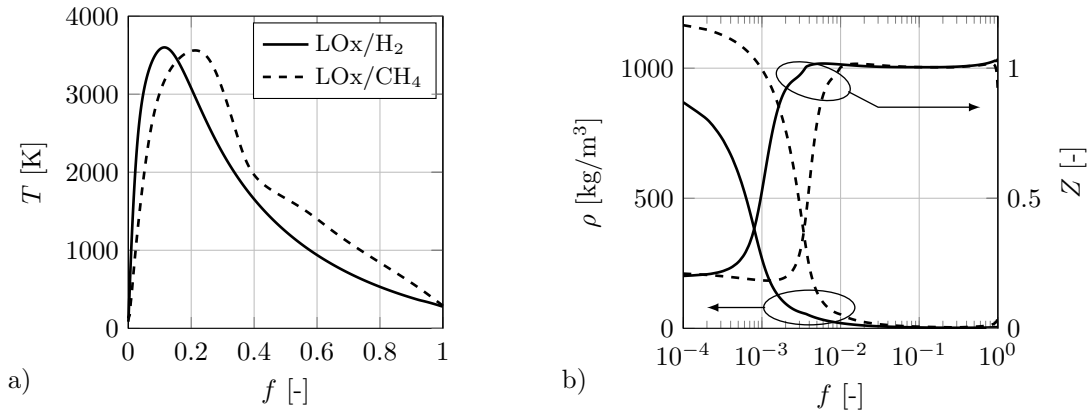

In Fig.~\ref{fig:flameletBase}~a the temperature profiles of the two flames are plotted over the mixture fraction~$f$. The maximum temperature is similar for both flames with different stoichiometric mixture fractions $f_{\textrm{st,LOx/H}_2}=0.055$ and $f_{\textrm{st,LOx/CH}_4}=0.2$, respectively. Consequently, the temperature gradient towards the oxidizer side of the hydrogen flamelet is steeper than the one of the methane flamelet. At the fuel-rich side, the temperature profiles have a similar shape but different absolute values at respective positions in the mixture fraction space. This results from the different stoichiometric mixture fractions and therefore the methane flame features larger temperatures at $f > 0.2$.\\

Figure~\ref{fig:flameletBase}~b shows the profiles of the density $\rho$ and the compressibility factor $Z$ of the two counterflow diffusion flames. Especially for $f < 0.01$ (at the oxdizer-rich side) significant changes for both properties can be seen and the compressibility factor is unequal to unity. Therefore, the thermodynamic states deviate from the ideal gas assumption ($Z = 1$) and real-gas effects are present. This can also be seen in the density profile which shows a strong decline over a small band in the mixture fraction space. In addition, the injection density in the LOx/CH$_{4}$ case is approximately 30\% higher compared to the LOx/H$_{2}$ case. This is the result of the different injection temperatures of oxygen which differ by 42~K between both flames, see Tab.~\ref{tab:operating conditions}. At the fuel-rich side real-gas effects ($Z \neq 1$) are also present. At the respective injection conditions methane has a compressibility factor of approximately 0.9 and hydrogen of 1.04. Due to the steady temperature increase for $1 > f > f_\textrm{st}$ the real-gas feature gets lost gradually and therefore $Z$ approximates unity over a wide part of the mixture fraction space. Most of the mixing and combustion takes therefore place under ideal-gas conditions. Summarizing, both sides of the flamelets show real-gas effects whereby the strongest deviation from the ideal gas assumption occurs at the oxidizer-rich side at $f < 0.01$. Similar findings have been reported by other groups~\cite{Ribert2008,Lacaze2012,banuti2016a,lapenna2017a}.

\subsubsection{Phase Separation}

\begin{figure}[!t]
	\centering
	\tikzsetnextfilename{flameletMascotte}%

\begin{tikzpicture}
	\begin{groupplot}[group style = {group name = plots,group size = 2 by 2, horizontal sep = 80pt, vertical sep = 70pt}, width = 0.34\textwidth, height = 0.29\textwidth, scale only axis]
	\nextgroupplot[
		xmajorgrids,
		ymajorgrids,
		xmin=0,
		xmax=1,
		xtick={ 0,0.2,0.4,0.6,0.8,1},
		xlabel={$f$ [-]},
		x tick label style={yshift=-2pt},
		ymin=0,
		ymax=4000,
		ytick={0, 1e3, 2e3, 3e3, 4e3},
		/pgf/number format/1000 sep={},
		ylabel={$T$ [K]},
		clip=false,
		legend style={font=\small,anchor=north east,legend cell align=left}
		]
		
		\addplot [color=black,solid,line width=1pt,smooth] table [y=T, x=x, col sep=comma]{../sources/Flamelets/Mascotte_GRI_RG_chi1_kij_0.3/postProcessing/sampleDict_thermo/0.00026562/thermo_T_ha_hs_rho_Cp_thermo:alpha_thermo:mu_HeatRelease_thermo:Z_thermo:psi_thermo:as_betaN.csv};
		
		
		\draw [color=gray,line width=1.0pt] (axis cs:-0.02,100) rectangle (axis cs:0.02,600) node (rectLeft) {};
		\draw[color=gray,line width=1.0pt] (rectLeft.west) -- (axis cs:0.67,2500);
		\draw [color=gray,line width=1.0pt] (axis cs:0.9,100) rectangle (axis cs:0.999,600) node (rectRight) {};
		\draw[color=gray,line width=1.0pt] (rectRight.west) -- (axis cs:0.67,2500);
		\node[draw=gray,fill=white,thick] at (axis cs:0.67,2500) [anchor=south,align=center] {\small $TPD < 0$ \\ \small $\Rightarrow$ phase separation};
	\nextgroupplot[
		xmajorgrids,
		ymajorgrids,
		xmin=0.0001,
		xmax=1,
		xlabel={$f$ [-]},
		x tick label style={yshift=-2pt},
		axis y line*=left,
		ymin=0,
		ymax=1,
		ylabel={$Y$ [-]},
		legend style={anchor=north east,legend cell align=left,align=left,draw=white!15!black}
		]
		
		\addplot [color=color1Pastel,solid,line width=1pt] table [y=CH4, x=x, col sep=comma]{../sources/Flamelets/Mascotte_GRI_RG_chi1_kij_0.3/postProcessing/sampleDict_species/0.00026562/species_N2_CH4_O2_CO_H2_H2O_CO2_OH_H_O.csv};\addlegendentry{CH$_4$}
		
		\addplot [color=color2Pastel,solid,line width=1pt] table [y=O2, x=x, col sep=comma]{../sources/Flamelets/Mascotte_GRI_RG_chi1_kij_0.3/postProcessing/sampleDict_species/0.00026562/species_N2_CH4_O2_CO_H2_H2O_CO2_OH_H_O.csv};\addlegendentry{O$_2$}
		
		\addplot [color=color3Pastel,solid,line width=1pt] table [y=H2O, x=x, col sep=comma]{../sources/Flamelets/Mascotte_GRI_RG_chi1_kij_0.3/postProcessing/sampleDict_species/0.00026562/species_N2_CH4_O2_CO_H2_H2O_CO2_OH_H_O.csv};\addlegendentry{H$_2$O}
		
		
	\nextgroupplot[
		xmode = log,
		xmajorgrids,
		ymajorgrids,
		xmin=0.0001,
		xmax=0.01,
		xlabel={$f$ [-]},
		x tick label style={yshift=-2pt},
		axis y line*=left,
		ymin=50,
		ymax=450,
		ylabel={$T$ [K]},
		colorbar horizontal,
		colorbar style={
			yshift = 202pt,
			ylabel={$\beta$ [-]},
			ylabel style={rotate=-90},
			x tick label style={
				/pgf/number format/.cd,
				fixed,
				fixed zerofill,
				precision=3,
				/tikz/.cd
			},
			font=\small
		}
		]
		
		\addplot [color=black,dashed,line width=1pt,smooth] table [y=T, x=x, col sep=comma]{../sources/Flamelets/Mascotte_GRI_RG_chi1_kij_0.3/postProcessing/sampleDict_thermo/0.00026222/thermo_T_ha_hs_rho_Cp_thermo:alpha_thermo:mu_HeatRelease_thermo:Z_thermo:psi_thermo:as.csv};
		\draw [color=black,line width=0.5pt,latex-] (axis cs:1.3e-4,90) -- (axis cs:1.6e-4,150) node[anchor=south west,draw=black,fill=white,align=center] {\scriptsize real-gas\\ \scriptsize reference};
		
		\addplot[only marks,fill=black] table [y=T, x=x, col sep=comma]{../sources/Flamelets/Mascotte_GRI_RG_chi1_kij_0.3/postProcessing/sampleDict_thermo/0.00026562/thermo_T_ha_hs_rho_Cp_thermo:alpha_thermo:mu_HeatRelease_thermo:Z_thermo:psi_thermo:as_betaN_1.csv};			
		\addplot 
		[		
		scatter/use mapped color=
		{
		    draw=black,
		    fill=mapped color
		},
		scatter,
		only marks, 
		point meta=explicit
		] table [y=T, x=x, meta=betaN, col sep=comma]{../sources/Flamelets/Mascotte_GRI_RG_chi1_kij_0.3/postProcessing/sampleDict_thermo/0.00026562/thermo_T_ha_hs_rho_Cp_thermo:alpha_thermo:mu_HeatRelease_thermo:Z_thermo:psi_thermo:as_betaN_2.csv};	
		\addplot[only marks,fill=black] table [y=T, x=x, col sep=comma]{../sources/Flamelets/Mascotte_GRI_RG_chi1_kij_0.3/postProcessing/sampleDict_thermo/0.00026562/thermo_T_ha_hs_rho_Cp_thermo:alpha_thermo:mu_HeatRelease_thermo:Z_thermo:psi_thermo:as_betaN_3.csv};	
		\node[anchor=north west,draw=black,fill=white,align=left] at (rel axis cs:0.02,0.98) {\small Detailed view: $f < 0.01$ \\ \rule{3.5cm}{0.4pt} \\ {\scriptsize black dots: single-phase}\\ {\scriptsize colored dots: two-phase}};
	\nextgroupplot[
		colorbar,
		xmajorgrids,
		ymajorgrids,
		xmin=0.9,
		xmax=1,
		xlabel={$f$ [-]},
		x tick label style={yshift=-2pt},
		axis y line*=left,
		ymin=100,
		ymax=600,
		ylabel={$T$ [K]},
		colorbar horizontal,
		colorbar style={
			yshift = 200pt,
			ylabel={$\beta$ [-]},
			ylabel style={rotate=-90},
			x tick label style={
				/pgf/number format/.cd,
				fixed,
				fixed zerofill,
				precision=4,
				/tikz/.cd
			},
			font=\small
		}
		]
		
		\addplot [color=black,dashed,line width=1pt,smooth] table [y=T, x=x, col sep=comma]{../sources/Flamelets/Mascotte_GRI_RG_chi1_kij_0.3/postProcessing/sampleDict_thermo/0.00026222/thermo_T_ha_hs_rho_Cp_thermo:alpha_thermo:mu_HeatRelease_thermo:Z_thermo:psi_thermo:as.csv};
		\draw [color=black,line width=0.5pt,latex-] (axis cs:0.957,420) -- (axis cs:0.961,435) node[anchor=south west,draw=black,fill=white,align=center] {\scriptsize real-gas\\ \scriptsize reference};	
		
		\addplot[only marks,fill=black] table [y=T, x=x, col sep=comma]{../sources/Flamelets/Mascotte_GRI_RG_chi1_kij_0.3/postProcessing/sampleDict_thermo/0.00026562/thermo_T_ha_hs_rho_Cp_thermo:alpha_thermo:mu_HeatRelease_thermo:Z_thermo:psi_thermo:as_betaN_3.csv};			
		\addplot 
		[		
		scatter/use mapped color=
		{
			draw=black,
			fill=mapped color
		},
		scatter,
		only marks, 
		point meta=explicit
		] table [y=T, x=x, meta=betaN, col sep=comma]{../sources/Flamelets/Mascotte_GRI_RG_chi1_kij_0.3/postProcessing/sampleDict_thermo/0.00026562/thermo_T_ha_hs_rho_Cp_thermo:alpha_thermo:mu_HeatRelease_thermo:Z_thermo:psi_thermo:as_betaN_4.csv};	
		\addplot[only marks,fill=black] table [y=T, x=x, col sep=comma]{../sources/Flamelets/Mascotte_GRI_RG_chi1_kij_0.3/postProcessing/sampleDict_thermo/0.00026562/thermo_T_ha_hs_rho_Cp_thermo:alpha_thermo:mu_HeatRelease_thermo:Z_thermo:psi_thermo:as_betaN_5.csv};
		\node[anchor=south west,draw=black,fill=white,align=left] at (rel axis cs:0.02,0.02) {\small Detailed view $0.9 < f < 1.0$ \\ \rule{4cm}{0.4pt} \\ {\scriptsize black dots: single-phase}\\ {\scriptsize colored dots: two-phase}};
	\end{groupplot}
   	\node[below = 0.4cm of plots c1r1.south,xshift=-4.0cm,yshift=-0.2cm] {a)};
   	\node[below = 0.4cm of plots c2r1.south,xshift=-4.0cm,yshift=-0.2cm] {b)};
   	\node[below = 0.4cm of plots c1r2.south,xshift=-4.0cm,yshift=-0.2cm] {c)};
   	\node[below = 0.4cm of plots c2r2.south,xshift=-4.0cm,yshift=-0.2cm] {d)};
\end{tikzpicture}%

	\caption{Thermodynamic analysis of the methane flamelet regarding phase separation effects: a) Temperature, b) Mass fractions for the three main components of the hydrogen flame, c) Detailed view of the oxidizer-rich side and d) Detailed view of the fuel-rich side.}
	\label{fig:flameletMascotte}
\end{figure}
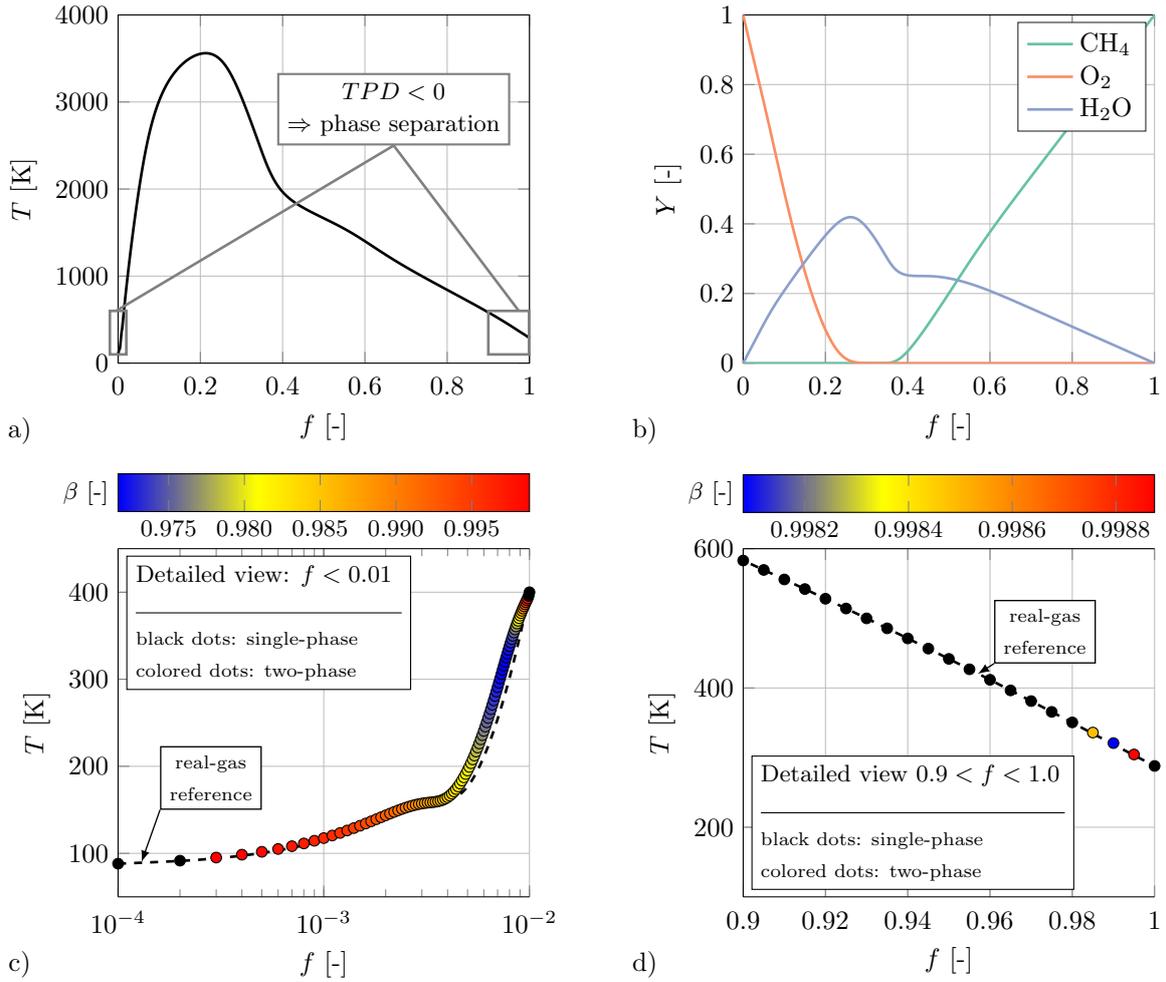

Figure~\ref{fig:flameletMascotte} shows the detailed analysis of the methane flame regarding phase separation. At both sides of the flamelet the stability analysis yields a negative tangent plane distance ($TPD < 0$), see Fig.~\ref{fig:flameletMascotte}~a, and therefore a phase separation was performed during the flamelet-calculation. The instability of the single-phase mixture can be attributed to moderate temperatures and the presence of water in these two regions, see Fig.~\ref{fig:flameletMascotte}~b. On both sides the presence of H$_2$O is caused by diffusion as the reaction rate of water is almost zero in these regions. In Fig.~\ref{fig:flameletMascotte}~c a detailed view for $T < 500$~K and $f < 0.01$ is shown by means of a scatter plot.
Black dots indicate a stable single-phase mixture and colored dots a two-phase mixture whereby the color corresponds to the vapor mole fraction $\beta$. As H$_2$O is the high volatile component in the mixture of the combustion products, see Fig.~\ref{fig:validationVLE}~c, and only a small amount of it is present in the mixture, see Fig.~\ref{fig:flameletMascotte}~b, $\beta$ is close to unity and ranges between 97\% and 100\%. Only three distinct points at the oxidizer side are stable within the mixture fraction space, namely $f=0$ (pure oxygen, not shown in Fig.~\ref{fig:flameletMascotte}~c), $f=1 \cdot 10^{-4}$ and $f=2 \cdot 10^{-4}$. All other points are unstable up to $f \approx 1 \cdot 10^{-2}$. Following the colored dots from left to right a continuous variation of $\beta$ can be seen although the composition of the mixture varies due to diffusion and production. At $f=3 \cdot 10^{-4}$ the dew-point line is crossed. Afterwards, the vapor mole fraction is reducing steadily until the minimum of approximately 0.97 is reached at $f \approx 8 \cdot 10^{-3}$. After the minimum is passed, $\beta$ rises and the dew-point line is crossed again at $f \approx 1 \cdot 10^{-2}$. This observation is very similar to the retrograde condensation discussed and illustrated in Fig.~\ref{fig:prudhoeBayVLE} for a fixed multicomponent mixture but here the retrograde character can be attributed to both the overall change in state as well as the characteristic of the multicomponent mixture. A similar behavior regarding the vapor fraction and the corresponding two-phase phenomena was reported by Traxinger~\textit{et~al.}~\cite{traxinger2017a} for a generic n-hexane nitrogen mixture at supercritical pressure. On the fuel-rich side a very similar pattern occurs but here the vapor mole fraction drops only to a minimum of about 0.998 at $f \approx 0.99$, see Fig.~\ref{fig:flameletMascotte}~d.\\

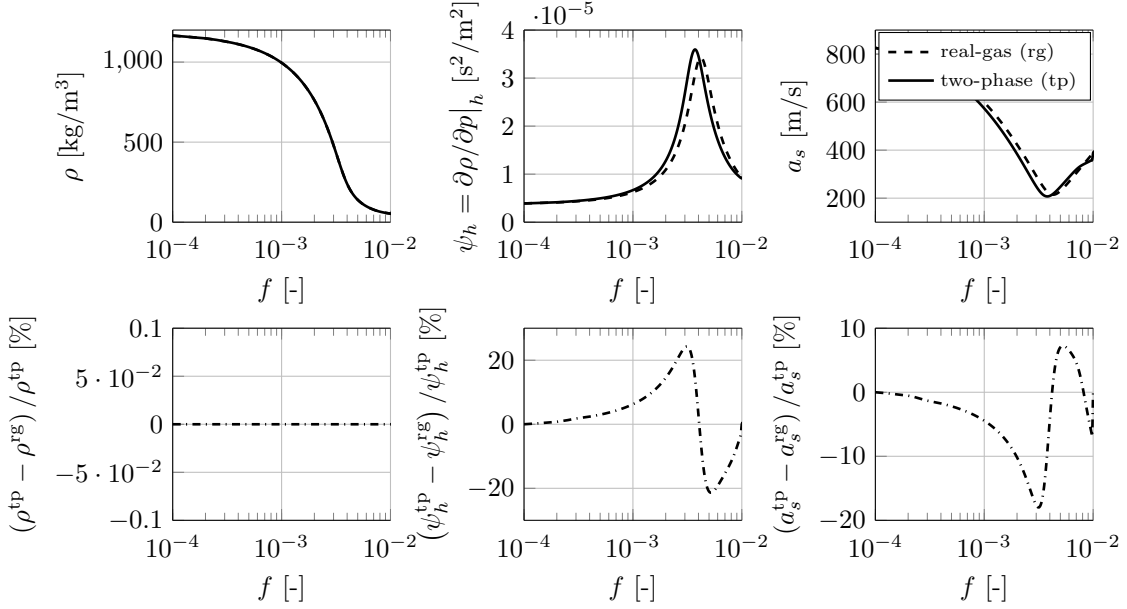
\begin{figure}[t]
	\centering
	\tikzsetnextfilename{flameletMascotte_diff}%

\begin{tikzpicture}
	\begin{groupplot}[group style = {group name = plots,group size = 3 by 2, horizontal sep = 50pt, vertical sep = 40pt}, width = 0.18\textwidth, height = 0.16\textwidth, scale only axis]	
	\nextgroupplot[
		xmode = log,
		xmajorgrids,
		ymajorgrids,
		xmin=0.0001,
		xmax=0.01,
		xlabel={$f$ [-]},
		x tick label style={yshift=-2pt},
		axis y line*=left,
		ymin=0,
		ymax=1200,
		ylabel={$\rho$ [kg/m$^3$]}
		]
		\addplot [color=black,dashed,line width=1pt,smooth] table [y=rho, x=x, col sep=comma]{../sources/Flamelets/Mascotte_GRI_RG_chi1_kij_0.3/postProcessing/sampleDict_thermo/0.00026222/thermo_T_ha_hs_rho_Cp_thermo:alpha_thermo:mu_HeatRelease_thermo:Z_thermo:psi_thermo:as.csv};
		\addplot [color=black,solid,line width=1pt,smooth] table [y=rho, x=x, col sep=comma]{../sources/Flamelets/Mascotte_GRI_RG_chi1_kij_0.3/postProcessing/sampleDict_thermo/0.00026562/thermo_T_ha_hs_rho_Cp_thermo:alpha_thermo:mu_HeatRelease_thermo:Z_thermo:psi_thermo:as_betaN.csv};
	\nextgroupplot[
		xmode = log,
		xmajorgrids,
		ymajorgrids,
		xmin=0.0001,
		xmax=0.01,
		xlabel={$f$ [-]},
		x tick label style={yshift=-2pt},
		axis y line*=left,
		ymin=0,
		ymax=4e-5,
		ylabel={$\psi_h = \partial \rho / \partial p \at[\big]{h} $ [s$^2$/m$^2$]}
		]
		\addplot [color=black,dashed,line width=1pt,smooth] table [y=thermo:psi, x=x, col sep=comma]{../sources/Flamelets/Mascotte_GRI_RG_chi1_kij_0.3/postProcessing/sampleDict_thermo/0.00026222/thermo_T_ha_hs_rho_Cp_thermo:alpha_thermo:mu_HeatRelease_thermo:Z_thermo:psi_thermo:as.csv};
		\addplot [color=black,solid,line width=1pt,smooth] table [y=thermo:psi, x=x, col sep=comma]{../sources/Flamelets/Mascotte_GRI_RG_chi1_kij_0.3/postProcessing/sampleDict_thermo/0.00026562/thermo_T_ha_hs_rho_Cp_thermo:alpha_thermo:mu_HeatRelease_thermo:Z_thermo:psi_thermo:as_betaN.csv};
	\nextgroupplot[
		xmode = log,
		xmajorgrids,
		ymajorgrids,
		xmin=0.0001,
		xmax=0.01,
		xlabel={$f$ [-]},
		x tick label style={yshift=-2pt},
		axis y line*=left,
		ymin=100,
		ymax=900,
		ylabel={$a_s$ [m/s]},
		legend style={font=\scriptsize,at={(rel axis cs: 0.99,0.99)},anchor=north east,legend cell align=left}
		]
		\addplot [color=black,dashed,line width=1pt,smooth] table [y=thermo:as, x=x, col sep=comma]{../sources/Flamelets/Mascotte_GRI_RG_chi1_kij_0.3/postProcessing/sampleDict_thermo/0.00026222/thermo_T_ha_hs_rho_Cp_thermo:alpha_thermo:mu_HeatRelease_thermo:Z_thermo:psi_thermo:as.csv};\addlegendentry{real-gas (rg)};
		\addplot [color=black,solid,line width=1pt,smooth] table [y=thermo:as, x=x, col sep=comma]{../sources/Flamelets/Mascotte_GRI_RG_chi1_kij_0.3/postProcessing/sampleDict_thermo/0.00026562/thermo_T_ha_hs_rho_Cp_thermo:alpha_thermo:mu_HeatRelease_thermo:Z_thermo:psi_thermo:as_betaN.csv};\addlegendentry{two-phase (tp)};
	\nextgroupplot[
		xmode = log,
		xmajorgrids,
		ymajorgrids,
		xmin=0.0001,
		xmax=0.01,
		xlabel={$f$ [-]},
		x tick label style={yshift=-2pt},
		axis y line*=left,
		ymin=-0.1,
		ymax=0.1,
		ylabel={$\left(\rho^\textrm{tp} - \rho^\textrm{rg}\right)/\rho^\textrm{tp}$ [\%]}
		]
		\addplot [color=black,dashdotted,line width=1pt,smooth] table [y=deltaRhoNorm, x=x, col sep=comma]{../sources/Flamelets/Mascotte_GRI_RG_chi1_kij_0.3/postProcessing/sampleDict_thermo/deltas.csv};
	\nextgroupplot[
		xmode = log,
		xmajorgrids,
		ymajorgrids,
		xmin=0.0001,
		xmax=0.01,
		xlabel={$f$ [-]},
		x tick label style={yshift=-2pt},
		axis y line*=left,
		ymin=-30,
		ymax=30,
		ylabel={$\left(\psi_h^\textrm{tp} - \psi_h^\textrm{rg}\right)/\psi_h^\textrm{tp}$ [\%]}
		]
		\addplot [color=black,dashdotted,line width=1pt,smooth] table [y=deltaPsiNorm, x=x, col sep=comma]{../sources/Flamelets/Mascotte_GRI_RG_chi1_kij_0.3/postProcessing/sampleDict_thermo/deltas.csv};
	\nextgroupplot[
		xmode = log,
		xmajorgrids,
		ymajorgrids,
		xmin=0.0001,
		xmax=0.01,
		xlabel={$f$ [-]},
		x tick label style={yshift=-2pt},
		axis y line*=left,
		ymin=-20,
		ymax=10,
		ylabel={$\left(a_s^\textrm{tp} - a_s^\textrm{rg}\right)/a_s^\textrm{tp}$ [\%]}
		]
		\addplot [color=black,dashdotted,line width=1pt,smooth] table [y=deltaAsNorm, x=x, col sep=comma]{../sources/Flamelets/Mascotte_GRI_RG_chi1_kij_0.3/postProcessing/sampleDict_thermo/deltas.csv};
	\end{groupplot}
\end{tikzpicture}%

	\caption{Comparison of the real-gas and the two-phase solution for the methane flamelet for $f < 0.01$.}
	\label{fig:flameletMascotte_diff}
\end{figure}

In both detailed views (Figs.~\ref{fig:flameletMascotte} c and d) the real-gas reference case (thermodynamic closure: "dense-gas" approach) is plotted as dashed line. The maximum difference in temperature occurs at the oxidizer-rich side and is approximately 40~K corresponding to a relative deviation of $\approx$14\%. For comparison, the maximum difference at the fuel-rich side is almost zero. Due to the consideration of the phase separation and hence the condensation of the fluid, the temperature of the two-phase case is larger, see Figs.~\ref{fig:flameletMascotte} c and d. Qiu and Reitz~\cite{qiu2015a} as well as Matheis and Hickel~\cite{matheis2018a} reported an identical behavior. The entrainment into the vapor liquid equilibrium region has also an effect onto other thermodynamic properties. In Fig.~\ref{fig:flameletMascotte_diff} the real-gas and the two-phase solution are compared to each other for $f<0.01$, where the largest deviations between both thermodynamic closures occur. For the density $\rho$ almost no deviation is found. 
For derived properties like the compressibility and the speed of sound things are different. As soon as the bubble- or dew-point line is crossed, these properties can be subject to large changes as the fluid is not homogeneous anymore, compare Fig.~\ref{fig:prudhoeBayAs}. In Fig.~\ref{fig:flameletMascotte_diff} the insenthalpic compressibility $\psi_h$ and the speed of sound $a_s$ of the real-gas and the two-phase flamelets are compared to each other. Both calculations show a similar shape for the respective properties. In the two-phase case, the speed of sound shows a drop at the points where the dew-point line is crossed. At the same position the compressibility has a slight discontinuity. Compared to the Prudhoe Bay mixture (see Fig.~\ref{fig:prudhoeBayAs}) and our previous investigation~\cite{traxinger2018a} the drop in speed of sound is minor and almost negligible for the methane flame. The direct comparison of the real-gas and the two-phase solution however shows deviations up to 20\%. This difference can mainly be attributed to the shift of the maximum in $\psi_h$ and the minimum in $a_s$ towards smaller mixture fractions in the two-phase solution. The main reason is the slightly higher temperature in the two-phase flamelet which is caused by the condensation of a small portion of gas indicated by $\beta$-values smaller than unity. Concluding, the consideration of the phase equilibrium leads to higher temperatures in the region with phase separation and therefore to the shift in the characteristic values which can be attributed to the pseudo-boiling process~\cite{oschwald2006a} in the real-gas case. Therefore, the solver stability is not significantly enhanced by considering the phase separation process as it was found for inert injection cases~\cite{traxinger2017a,traxinger2018a,matheis2018a}. The only phenomenon, which might have a significant effect on the solver stability is the contribution of the surface tension inside the momentum equation. This effect is  neglected within this study.\\

In Figs.~\ref{fig:flameletBKH_diff} and \ref{fig:flameletBKH} the phase separation effects occurring in the hydrogen counterflow diffusion flame are presented. The solution of the hydrogen flamelet leads to similar findings as the thoroughly discussed methane flamelet. Instable single-phase solutions occur at both sides of the mixture fraction space and have similar effects on the solution and the derived thermodynamic properties. Due to the significant temperature differences at already small mixture fractions, the deviation in the calculated density ranges up to almost 40\%. This is a huge difference compared to the methane flamelet. The conclusions regarding the shape and the shift discussed for the LOx/CH$_4$ case remain the same.

\clearpage

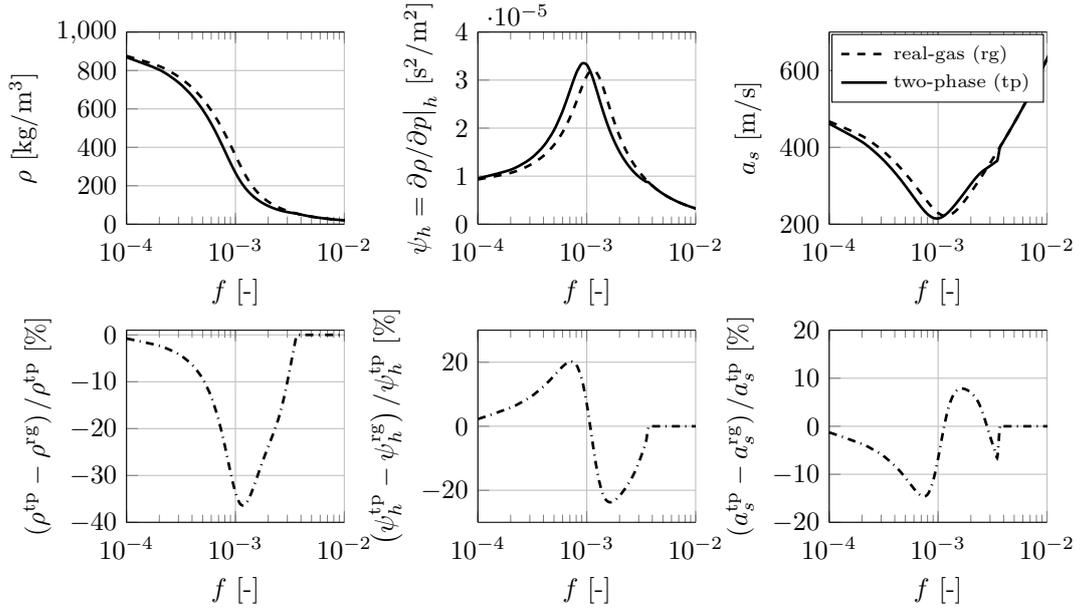
\begin{figure}[!h]
	\centering
	\tikzsetnextfilename{flameletBKH_diff}%

\begin{tikzpicture}
	\begin{groupplot}[group style = {group name = plots,group size = 3 by 2, horizontal sep = 50pt, vertical sep = 40pt}, width = 0.18\textwidth, height = 0.16\textwidth, scale only axis]	
	\nextgroupplot[
		xmode = log,
		xmajorgrids,
		ymajorgrids,
		xmin=0.0001,
		xmax=0.01,
		xlabel={$f$ [-]},
		x tick label style={yshift=-2pt},
		axis y line*=left,
		ymin=0,
		ymax=1000,
		ylabel={$\rho$ [kg/m$^3$]}
		]
		\addplot [color=black,dashed,line width=1pt,smooth] table [y=rho, x=x, col sep=comma]{../sources/Flamelets/BKH_Connaire_chi1_kij_0.3/postProcessing/sampleDict_thermo/0.01235501/thermo_T_ha_hs_rho_Cp_thermo:alpha_thermo:mu_HeatRelease_thermo:Z_thermo:psi_thermo:as.csv};
		\addplot [color=black,solid,line width=1pt,smooth] table [y=rho, x=x, col sep=comma]{../sources/Flamelets/BKH_Connaire_chi1_kij_0.3/postProcessing/sampleDict_thermo/0.01240101/thermo_T_ha_hs_rho_Cp_thermo:alpha_thermo:mu_HeatRelease_thermo:Z_thermo:psi_thermo:as_betaN.csv};
	\nextgroupplot[
		xmode = log,
		xmajorgrids,
		ymajorgrids,
		xmin=0.0001,
		xmax=0.01,
		xlabel={$f$ [-]},
		x tick label style={yshift=-2pt},
		axis y line*=left,
		ymin=0,
		ymax=4e-5,
		ylabel={$\psi_h = \partial \rho / \partial p \at[\big]{h} $ [s$^2$/m$^2$]}
		]
		\addplot [color=black,dashed,line width=1pt,smooth] table [y=thermo:psi, x=x, col sep=comma]{../sources/Flamelets/BKH_Connaire_chi1_kij_0.3/postProcessing/sampleDict_thermo/0.01235501/thermo_T_ha_hs_rho_Cp_thermo:alpha_thermo:mu_HeatRelease_thermo:Z_thermo:psi_thermo:as.csv};
		\addplot [color=black,solid,line width=1pt,smooth] table [y=thermo:psi, x=x, col sep=comma]{../sources/Flamelets/BKH_Connaire_chi1_kij_0.3/postProcessing/sampleDict_thermo/0.01240101/thermo_T_ha_hs_rho_Cp_thermo:alpha_thermo:mu_HeatRelease_thermo:Z_thermo:psi_thermo:as_betaN.csv};
	\nextgroupplot[
		xmode = log,
		xmajorgrids,
		ymajorgrids,
		xmin=0.0001,
		xmax=0.01,
		xlabel={$f$ [-]},
		x tick label style={yshift=-2pt},
		axis y line*=left,
		ymin=200,
		ymax=700,
		ylabel={$a_s$ [m/s]},
		legend style={font=\scriptsize,at={(rel axis cs: 0.01,0.99)},anchor=north west,legend cell align=left}
		]
		\addplot [color=black,dashed,line width=1pt,smooth] table [y=thermo:as, x=x, col sep=comma]{../sources/Flamelets/BKH_Connaire_chi1_kij_0.3/postProcessing/sampleDict_thermo/0.01235501/thermo_T_ha_hs_rho_Cp_thermo:alpha_thermo:mu_HeatRelease_thermo:Z_thermo:psi_thermo:as.csv};\addlegendentry{real-gas (rg)};
		\addplot [color=black,solid,line width=1pt,smooth] table [y=thermo:as, x=x, col sep=comma]{../sources/Flamelets/BKH_Connaire_chi1_kij_0.3/postProcessing/sampleDict_thermo/0.01240101/thermo_T_ha_hs_rho_Cp_thermo:alpha_thermo:mu_HeatRelease_thermo:Z_thermo:psi_thermo:as_betaN.csv};\addlegendentry{two-phase (tp)};
	\nextgroupplot[
		xmode = log,
		xmajorgrids,
		ymajorgrids,
		xmin=0.0001,
		xmax=0.01,
		xlabel={$f$ [-]},
		x tick label style={yshift=-2pt},
		axis y line*=left,
		ymin=-40,
		ymax=1,
		ylabel={$\left(\rho^\textrm{tp} - \rho^\textrm{rg}\right)/\rho^\textrm{tp}$ [\%]}
		]
		\addplot [color=black,dashdotted,line width=1pt,smooth] table [y=deltaRhoNorm, x=x, col sep=comma]{../sources/Flamelets/BKH_Connaire_chi1_kij_0.3/postProcessing/sampleDict_thermo/deltas.csv};
	\nextgroupplot[
		xmode = log,
		xmajorgrids,
		ymajorgrids,
		xmin=0.0001,
		xmax=0.01,
		xlabel={$f$ [-]},
		x tick label style={yshift=-2pt},
		axis y line*=left,
		ymin=-30,
		ymax=30,
		ylabel={$\left(\psi_h^\textrm{tp} - \psi_h^\textrm{rg}\right)/\psi_h^\textrm{tp}$ [\%]}
		]
		\addplot [color=black,dashdotted,line width=1pt,smooth] table [y=deltaPsiNorm, x=x, col sep=comma]{../sources/Flamelets/BKH_Connaire_chi1_kij_0.3/postProcessing/sampleDict_thermo/deltas.csv};
	\nextgroupplot[
		xmode = log,
		xmajorgrids,
		ymajorgrids,
		xmin=0.0001,
		xmax=0.01,
		xlabel={$f$ [-]},
		x tick label style={yshift=-2pt},
		axis y line*=left,
		ymin=-20,
		ymax=20,
		ylabel={$\left(a_s^\textrm{tp} - a_s^\textrm{rg}\right)/a_s^\textrm{tp}$ [\%]}
		]
		\addplot [color=black,dashdotted,line width=1pt,smooth] table [y=deltaAsNorm, x=x, col sep=comma]{../sources/Flamelets/BKH_Connaire_chi1_kij_0.3/postProcessing/sampleDict_thermo/deltas.csv};
	\end{groupplot}
\end{tikzpicture}%

	\vspace*{-6pt}
	\caption{Comparison of the real-gas and the two-phase solution for the hydrogen flamelet for $f < 0.01$.}
	\label{fig:flameletBKH_diff}
	\vspace*{-6pt}
\end{figure}
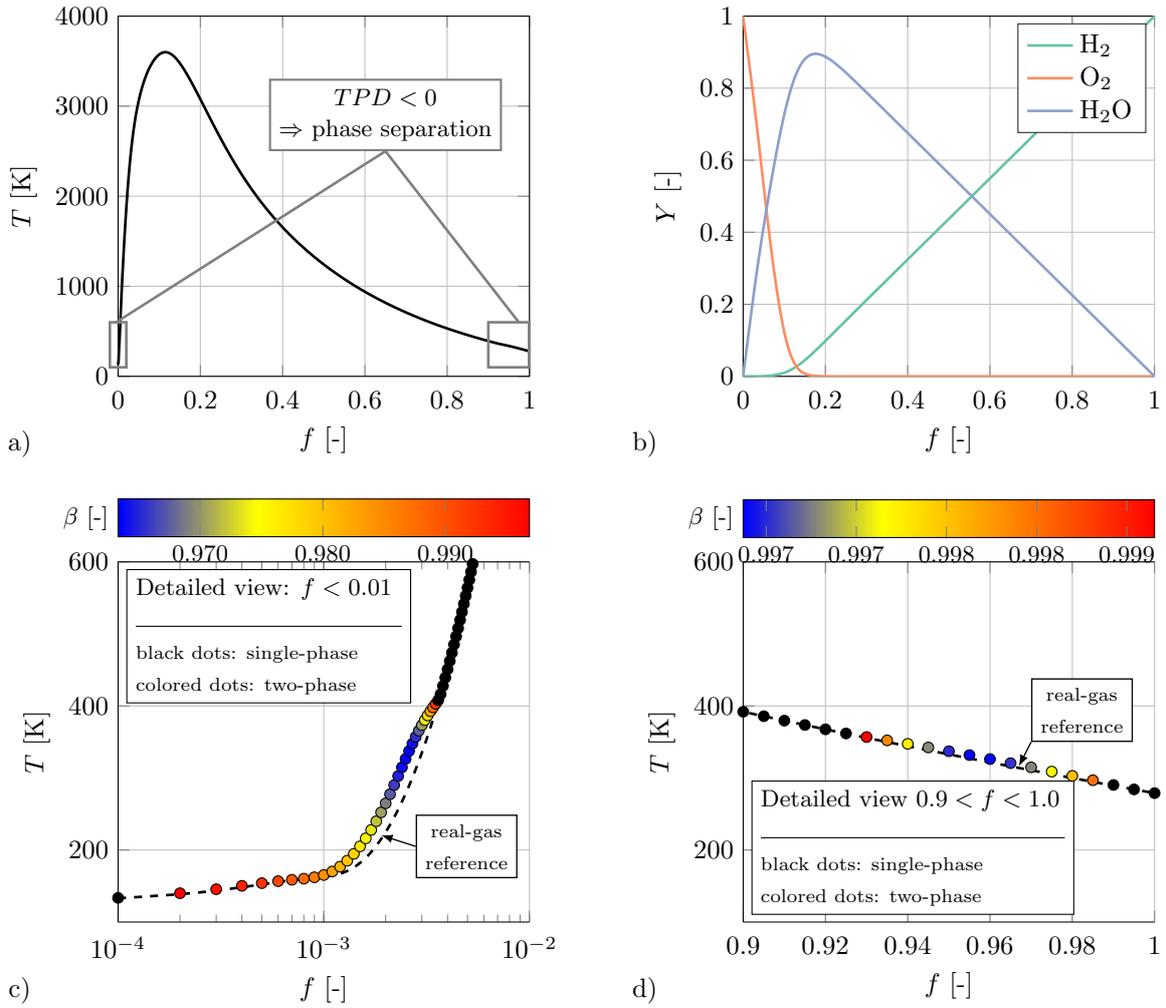
\begin{figure}[!h]
	\centering
	\tikzsetnextfilename{flameletBKH}%

\begin{tikzpicture}
	\begin{groupplot}[group style = {group name = plots,group size = 2 by 2, horizontal sep = 80pt, vertical sep = 70pt}, width = 0.34\textwidth, height = 0.30\textwidth, scale only axis]
	\nextgroupplot[
		xmajorgrids,
		ymajorgrids,
		xmin=0,
		xmax=1,
		xtick={ 0,0.2,0.4,0.6,0.8,1},
		xlabel={$f$ [-]},
		x tick label style={yshift=-2pt},
		ymin=0,
		ymax=4000,
		ytick={0, 1e3, 2e3, 3e3, 4e3},
		/pgf/number format/1000 sep={},
		ylabel={$T$ [K]},
		clip=false,
		legend style={font=\small,anchor=north east,legend cell align=left}
		]
		
		\addplot [color=black,solid,line width=1pt,smooth] table [y=T, x=x, col sep=comma]{../sources/Flamelets/BKH_Connaire_chi1_kij_0.3/postProcessing/sampleDict_thermo/0.01240101/thermo_T_ha_hs_rho_Cp_thermo:alpha_thermo:mu_HeatRelease_thermo:Z_thermo:psi_thermo:as_betaN.csv};
		
		
		\draw [color=gray,line width=1.0pt] (axis cs:-0.02,100) rectangle (axis cs:0.02,600) node (rectLeft) {};
		\draw[color=gray,line width=1.0pt] (rectLeft.west) -- (axis cs:0.65,2500);
		\draw [color=gray,line width=1.0pt] (axis cs:0.9,100) rectangle (axis cs:0.999,600) node (rectRight) {};
		\draw[color=gray,line width=1.0pt] (rectRight.west) -- (axis cs:0.65,2500);
		\node[draw=gray,fill=white,thick] at (axis cs:0.65,2500) [anchor=south,align=center] {\small $TPD < 0$ \\ \small $\Rightarrow$ phase separation};
	\nextgroupplot[
		xmajorgrids,
		ymajorgrids,
		xmin=0.0001,
		xmax=1,
		xlabel={$f$ [-]},
		x tick label style={yshift=-2pt},
		axis y line*=left,
		ymin=0,
		ymax=1,
		ylabel={$Y$ [-]},
		legend style={anchor=north east,legend cell align=left,align=left,draw=white!15!black}
		]
		
		\addplot [color=color1Pastel,solid,line width=1pt] table [y=H2, x=x, col sep=comma]{../sources/Flamelets/BKH_Connaire_chi1_kij_0.3/postProcessing/sampleDict_species/0.01240101/species_H_H2_O_O2_OH_H2O_HO2_H2O2.csv};\addlegendentry{H$_2$}
		
		\addplot [color=color2Pastel,solid,line width=1pt] table [y=O2, x=x, col sep=comma]{../sources/Flamelets/BKH_Connaire_chi1_kij_0.3/postProcessing/sampleDict_species/0.01240101/species_H_H2_O_O2_OH_H2O_HO2_H2O2.csv};\addlegendentry{O$_2$}
		
		\addplot [color=color3Pastel,solid,line width=1pt] table [y=H2O, x=x, col sep=comma]{../sources/Flamelets/BKH_Connaire_chi1_kij_0.3/postProcessing/sampleDict_species/0.01240101/species_H_H2_O_O2_OH_H2O_HO2_H2O2.csv};\addlegendentry{H$_2$O}
		
	\nextgroupplot[
		xmode = log,
		xmajorgrids,
		ymajorgrids,
		xmin=0.0001,
		xmax=0.01,
		xlabel={$f$ [-]},
		x tick label style={yshift=-2pt},
		axis y line*=left,
		ymin=100,
		ymax=600,
		ylabel={$T$ [K]},
		colorbar horizontal,
		colorbar style={
			yshift = 202pt,
			ylabel={$\beta$ [-]},
			ylabel style={rotate=-90},
			x tick label style={
				/pgf/number format/.cd,
				fixed,
				fixed zerofill,
				precision=3,
				/tikz/.cd
			},
			font=\small
		}
		]
		\addplot [color=black,dashed,line width=1pt,smooth] table [y=T, x=x, col sep=comma]{../sources/Flamelets/BKH_Connaire_chi1_kij_0.3/postProcessing/sampleDict_thermo/0.01235501/thermo_T_ha_hs_rho_Cp_thermo:alpha_thermo:mu_HeatRelease_thermo:Z_thermo:psi_thermo:as.csv};
		\draw [color=black,line width=0.5pt,latex-] (axis cs:0.0019,220.53) -- (axis cs:0.0028,205) node[anchor=west,draw=black,fill=white,align=center] {\scriptsize real-gas\\ \scriptsize reference};
		\addplot[only marks,fill=black] table [y=T, x=x, col sep=comma]{../sources/Flamelets/BKH_Connaire_chi1_kij_0.3/postProcessing/sampleDict_thermo/0.01240101/thermo_T_ha_hs_rho_Cp_thermo:alpha_thermo:mu_HeatRelease_thermo:Z_thermo:psi_thermo:as_betaN_1.csv};			
		\addplot 
		[		
		scatter/use mapped color=
		{
		    draw=black,
		    fill=mapped color
		},
		scatter,
		only marks, 
		point meta=explicit
		] table [y=T, x=x, meta=betaN, col sep=comma]{../sources/Flamelets/BKH_Connaire_chi1_kij_0.3/postProcessing/sampleDict_thermo/0.01240101/thermo_T_ha_hs_rho_Cp_thermo:alpha_thermo:mu_HeatRelease_thermo:Z_thermo:psi_thermo:as_betaN_2.csv};	
		\addplot[only marks,fill=black] table [y=T, x=x, col sep=comma]{../sources/Flamelets/BKH_Connaire_chi1_kij_0.3/postProcessing/sampleDict_thermo/0.01240101/thermo_T_ha_hs_rho_Cp_thermo:alpha_thermo:mu_HeatRelease_thermo:Z_thermo:psi_thermo:as_betaN_3.csv};	
		\node[anchor=north west,draw=black,fill=white,align=left] at (rel axis cs:0.02,0.98) {\small Detailed view: $f < 0.01$ \\ \rule{3.5cm}{0.4pt} \\ {\scriptsize black dots: single-phase}\\ {\scriptsize colored dots: two-phase}};
	\nextgroupplot[
		colorbar,
		xmajorgrids,
		ymajorgrids,
		xmin=0.9,
		xmax=1,
		xlabel={$f$ [-]},
		x tick label style={yshift=-2pt},
		axis y line*=left,
		ymin=100,
		ymax=600,
		ylabel={$T$ [K]},
		colorbar horizontal,
		colorbar style={
			yshift = 200pt,
			ylabel={$\beta$ [-]},
			ylabel style={rotate=-90},
			x tick label style={
				/pgf/number format/.cd,
				fixed,
				fixed zerofill,
				precision=3,
				/tikz/.cd
			},
			font=\small
		}
		]
		\addplot [color=black,dashed,line width=1pt,smooth] table [y=T, x=x, col sep=comma]{../sources/Flamelets/BKH_Connaire_chi1_kij_0.3/postProcessing/sampleDict_thermo/0.01235501/thermo_T_ha_hs_rho_Cp_thermo:alpha_thermo:mu_HeatRelease_thermo:Z_thermo:psi_thermo:as.csv};	
		\draw [color=black,line width=0.5pt,latex-] (axis cs:0.967,316.12) -- (axis cs:0.97,350) node[anchor=south west,draw=black,fill=white,align=center] {\scriptsize real-gas\\ \scriptsize reference};	
		\addplot[only marks,fill=black] table [y=T, x=x, col sep=comma]{../sources/Flamelets/BKH_Connaire_chi1_kij_0.3/postProcessing/sampleDict_thermo/0.01240101/thermo_T_ha_hs_rho_Cp_thermo:alpha_thermo:mu_HeatRelease_thermo:Z_thermo:psi_thermo:as_betaN_3.csv};			
		\addplot 
		[		
		scatter/use mapped color=
		{
			draw=black,
			fill=mapped color
		},
		scatter,
		only marks, 
		point meta=explicit
		] table [y=T, x=x, meta=betaN, col sep=comma]{../sources/Flamelets/BKH_Connaire_chi1_kij_0.3/postProcessing/sampleDict_thermo/0.01240101/thermo_T_ha_hs_rho_Cp_thermo:alpha_thermo:mu_HeatRelease_thermo:Z_thermo:psi_thermo:as_betaN_4.csv};	
		\addplot[only marks,fill=black] table [y=T, x=x, col sep=comma]{../sources/Flamelets/BKH_Connaire_chi1_kij_0.3/postProcessing/sampleDict_thermo/0.01240101/thermo_T_ha_hs_rho_Cp_thermo:alpha_thermo:mu_HeatRelease_thermo:Z_thermo:psi_thermo:as_betaN_5.csv};
		\node[anchor=south west,draw=black,fill=white,align=left] at (rel axis cs:0.02,0.02) {\small Detailed view $0.9 < f < 1.0$ \\ \rule{4cm}{0.4pt} \\ {\scriptsize black dots: single-phase}\\ {\scriptsize colored dots: two-phase}};
	\end{groupplot}
   	\node[below = 0.4cm of plots c1r1.south,xshift=-4.0cm,yshift=-0.2cm] {a)};
   	\node[below = 0.4cm of plots c2r1.south,xshift=-4.0cm,yshift=-0.2cm] {b)};
   	\node[below = 0.4cm of plots c1r2.south,xshift=-4.0cm,yshift=-0.2cm] {c)};
   	\node[below = 0.4cm of plots c2r2.south,xshift=-4.0cm,yshift=-0.2cm] {d)};
\end{tikzpicture}%

	\caption{Thermodynamic analysis of the hydrogen flamelet regarding phase separation effects: a) Temperature, b) Mass fractions for the three main components of the methane flame, c) Detailed view of the oxidizer-rich side and d) Detailed view of the fuel-rich side.}
	\label{fig:flameletBKH}
	\vspace*{-12pt}
\end{figure}
\subsection{Test Case and Numerical Setup} \label{subsec: Test Cases and Setup}

For the evaluation of the role of real-gas and two-phase phenomena under rocket-like conditions, the experiment of Singla \textit{et al.}~\cite{Singla2005} conducted at the Mascotte test bench is chosen as a reference. This test case has been previously used as validation case for numerical simulations, see, e.g., Schmitt \textit{et al.}~\cite{Schmitt2011}, Kim \textit{et al.}~\cite{Kim2013} or Cutrone~\textit{et al.}~\cite{Cutrone2010}. At an operating pressure of 56.1~bar, oxygen is injected in a liquid-like state at 85~K and the injection temperature of methane is 288~K, see Tab.~\ref{tab:operating conditions}. The combustion chamber is a 50~mm $\times$ 50~mm square duct with a length of 400~mm and a convergent-divergent nozzle. Being a single-element test case, the propellants are supplied by one coaxial injection element with 3.4~mm LOx pipe diameter, a CH$_4$ annulus height of 2.2~mm and a posttip width of 0.6~mm. As windows allow for optical access, Singla \textit{et al.}~\cite{Singla2005} measured spontaneous emission of OH$^\ast$ and CH$^\ast$. Besides the line-of-sight integrated results they employed an Abel transformation to obtain slices through the flame which can be additionally used for the comparison with the LES results. 
\begin{figure}[!h]
	\begin{center}
		\input{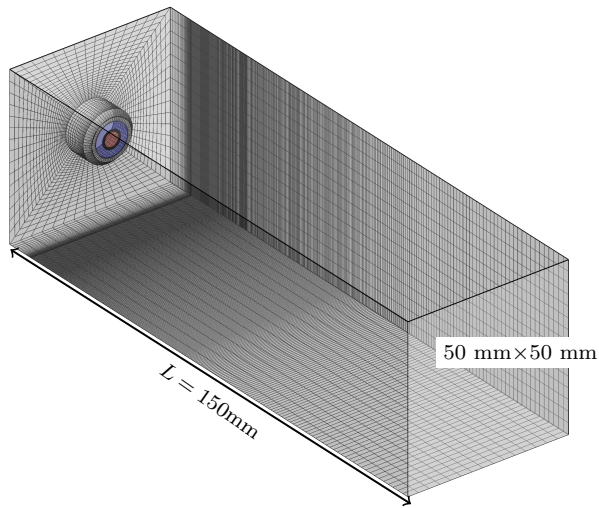}
	\end{center}
	\vspace*{0pt}
	\caption{Computational domain of the LOx/CH$_4$ test case.}
	\label{fig: Mascotte Domain}
\end{figure}
The computational domain (see Fig.~\ref{fig: Mascotte Domain}) is truncated after 150~mm, which is sufficient to accommodate the flame and to avoid interactions with the outlet boundary. All walls are considered as adiabatic no-slip boundary conditions. Realistic turbulent inflow boundary conditions for LOx and CH$_4$ jets are generated using separate incompressible LES at corresponding operating conditions with cyclic boundaries in axial direction. The turbulent velocity field of those precursor LES is extracted and interpolated in space and time onto the grid of the actual simulation. The reference grid consists of $16~\cdot 10^6$ cells with a minimum resolution at the injector of 0.2~mm in axial and 0.025~mm in radial direction. Figure~\ref{fig: Mascotte Domain} schematically shows the domain and the qualitative mesh resolution with reduced grid density.
\subsection{LES Results} \label{subsec: Results}

\subsubsection{General Flame Features}
\begin{figure}[t]
  \centering
   \input{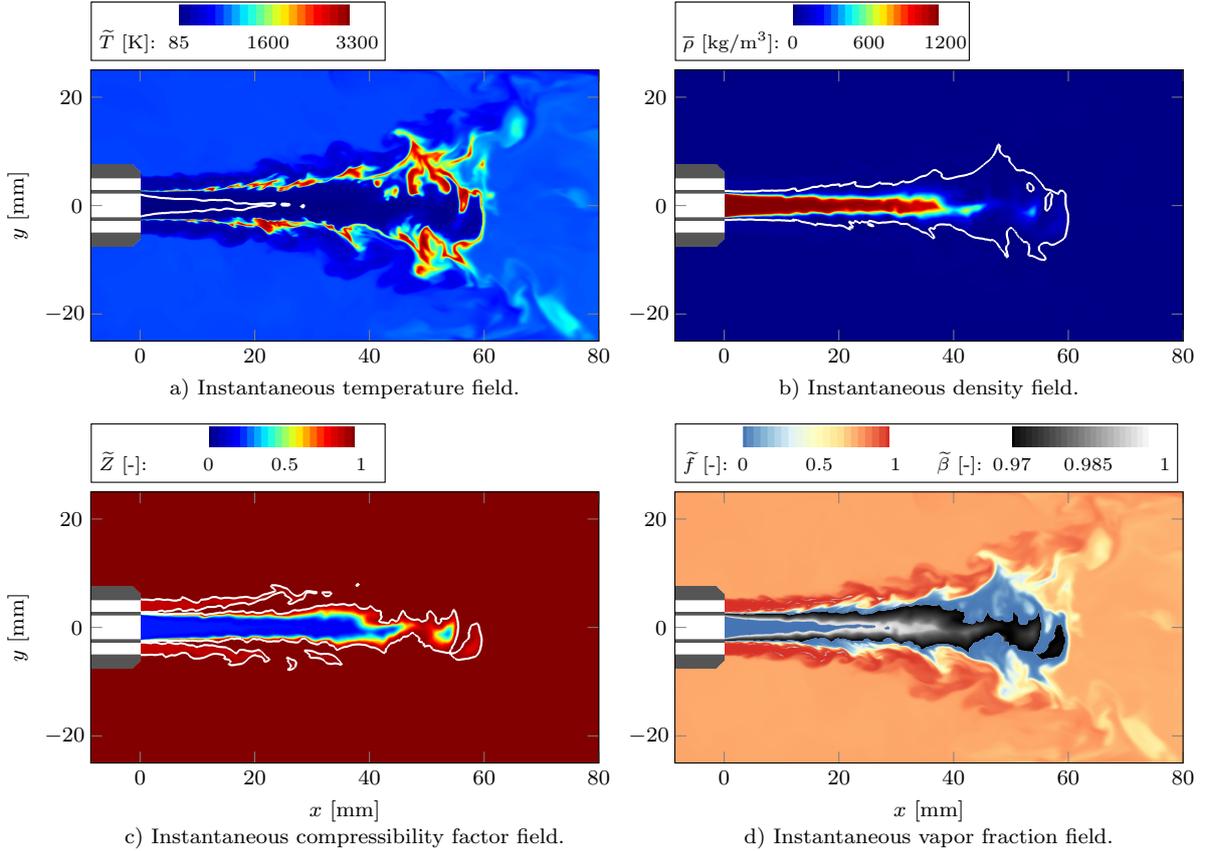}
  \caption{Instantaneous view of the $x$-$y$ plane of the LOx/CH$_4$ flame: The white line in a) encloses the region of $f \leq 2\cdot10^{-4}$ and in b) it represents the stoichiometric mixture fraction $f_{st}=0.2$. In c) the white line is drawn for $Z = 0.99$. In d) the $\beta$-field is superimposed onto the mixture fraction field $f$.}
  \label{fig: Mascotte general flame features}
\end{figure}

Figure~\ref{fig: Mascotte general flame features} shows various instantaneous fields of the LOx/CH$_4$ flame using a central cut through the flame to provide a qualitative impression of the flame shape. The instantaneous temperature field, see Fig.~\ref{fig: Mascotte general flame features}~a, indicates that a thin diffusion flame emanates from the injector which is anchored at the posttip. As the flame is submitted to a high level of strain due to the strong velocity gradient between oxidizer and fuel stream, the surface is affected by small turbulent structures. The instabilities increase further downstream and lead to larger turbulent structures and a strongly wrinkled flame. Chemical conversion leads to a radial expansion and it can be seen that the region of highest temperature ends at $x \approx 60$~mm, where the oxidizer is fully consumed. The white line in Fig.~\ref{fig: Mascotte general flame features}~a indicates the mixture fraction $f = 2\cdot10^{-4}$ and encloses therefore the region of (almost) pure oxygen. In Fig.~\ref{fig: Mascotte general flame features}~b the instantaneous density field is shown. Disintegration of the stable cold oxygen core starts at $x \approx 40$~mm, where pockets of high-density fluid detach from the core and are transported downstream. There, they react with the surrounding fuel leading to the region of strongest chemical reactions. The white line in Fig.~\ref{fig: Mascotte general flame features}~b denotes the stoichiometric mixture fraction $f_{st}=0.2$ highlighting the position of the flame.
As elaborated by Schmitt \textit{et al.}~\cite{Schmitt2011} using their LES data, the characteristic flame shape results from the interaction between flow and chamber wall. Besides the typical recirculation zones of single-element combustors between fuel jet and wall, a second large recirculation zone exists in the middle of the chamber between $x \approx 60$~mm and $x \approx 120$~mm, i.e., downstream of the main reaction region. As the hot gases are radially accelerated, they interact with the wall and are consequently redirected towards the center of the combustion chamber. Here, the combustion products flow towards the reaction zone and thereby increase the expansion of the flame, finally forming a stable condition. In order to emphasis the real-gas character of the investigated flame, Fig.~\ref{fig: Mascotte general flame features}~c shows the compressibility factor $Z$. The scale is truncated at unity and the region of  real-gas behavior is highlighted by the white iso-line drawn for $Z = 0.99$. As already discussed based on Fig.~\ref{fig:flameletMascotte}, both the oxidizer as well as the fuel feature real-gas effects whereby the strongest deviation from the ideal gas state are present on the oxygen side.

\subsubsection{Phase-Separation Effects}

Figure~\ref{fig: Mascotte general flame features}~d shows the instantaneous vapor fraction field $\beta$ superimposed onto the mixture fraction $f$ which is used to indicate the general flame structure and shape. In accordance to Fig.~\ref{fig:flameletMascotte} $\beta$ ranges between 97\% and 100\%. The largest region with phase separation is present within the LOx-core where combustion products and especially water mixes with the pure oxygen. Only for pure oxygen no phase instability was found as the injection and combustion is almost isobaric and therefore supercritical with respect to the critical point of oxygen. Minor areas of phase separation are also present around the pure fuel. In Fig.~\ref{fig: Mascotte general flame features}~d these regions are almost not visible, but Fig.~\ref{fig:flameletMascotte}~d proves their presence as we used the counter-flow diffusion flame concept to build up our thermo-chemical library for the LES. Furthermore, the strain-rate and hence the deviation of the scalar dissipation rate $\chi$ from unity is very small.

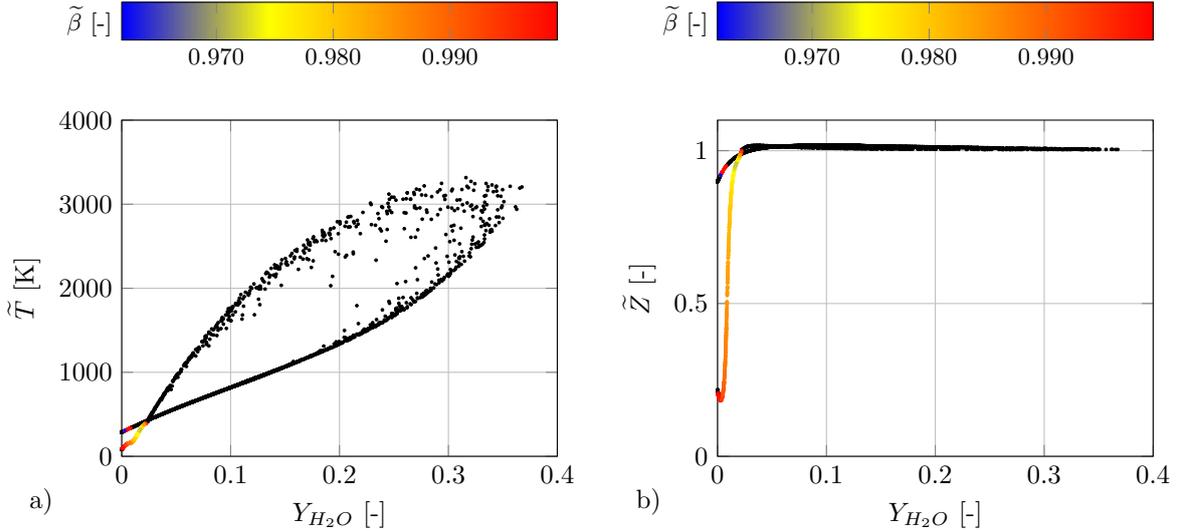
\begin{figure}[t]
  \centering
   \tikzsetnextfilename{Mascotte_scatter_midplane}

\begin{tikzpicture}
\begin{groupplot}[group style = {group size = 2 by 1, vertical sep=1cm, horizontal sep=60pt }, width=0.36\textwidth,height=0.28\textwidth]
	\nextgroupplot[%
	xmajorgrids,
	ymajorgrids,
	scale only axis,
	xmin=0,
	xmax=0.4,
	xtick={ 0,0.1,  0.2, 0.3, 0.4},
	xlabel={$Y_{H_2O}$ [-]},
	ymin=0,
	ymax=4000,
	ymajorgrids,
	ytick={0, 1e3, 2e3, 3e3, 4e3},
	ylabel={$\widetilde{T}$ [K]},
	/pgf/number format/1000 sep={},
	ylabel style={yshift=-0.0cm},
	legend style={font=\footnotesize,at={(0,1)},anchor=north west,legend cell align=left,align=left,draw=white!15!black, xshift=1pt, yshift=-1pt},
	colorbar horizontal,
	colorbar style={
		yshift = 210pt,
		ylabel={$\widetilde{\beta}$ [-]},
		ylabel style={rotate=-90},
		x tick label style={
			/pgf/number format/.cd,
			fixed,
			fixed zerofill,
			precision=3,
			/tikz/.cd
		},
		font=\small
	}
	]
	
	\addplot [only marks,mark size=0.5pt,fill=black,each nth point=9] table [x=H2O,y=T, col sep=comma]{../sources/Mascotte/Mascotte_scatter_midplane_betaLess095.csv};
	\addplot 
	[		
	scatter/use mapped color=
	{
	draw opacity=0,
	fill=mapped color
	},
	scatter,
	only marks, 
	mark size=0.7pt,
	point meta=explicit,
	each nth point=8
	] table [x=H2O,y=T, meta=beta, col sep=comma]{../sources/Mascotte/Mascotte_scatter_midplane_betaMore095.csv};
	\nextgroupplot[%
	xmajorgrids,
	ymajorgrids,
	scale only axis,
	xmin=0,
	xmax=0.4,
	xtick={ 0,0.1,  0.2, 0.3, 0.4},
	xlabel={$Y_{H_2O}$ [-]},
	ymin=0,
	ymax=1.1,
	ymajorgrids,
	ytick={0, 0.5, 1},
	ylabel={$\widetilde{Z}$ [-]},
	colorbar horizontal,
	colorbar style={
		yshift = 210pt,
		ylabel={$\widetilde{\beta}$ [-]},
		ylabel style={rotate=-90},
		x tick label style={
			/pgf/number format/.cd,
			fixed,
			fixed zerofill,
			precision=3,
			/tikz/.cd
		},
		font=\small
	}
	]
	
	\addplot [only marks,mark size=0.5pt,fill=black,each nth point=9] table [x=H2O,y=Z, col sep=comma]{../sources/Mascotte/Mascotte_scatter_midplane_betaLess095.csv};
	\addplot 
	[		
	scatter/use mapped color=
	{
	draw opacity=0,
	fill=mapped color
	},
	scatter,
	only marks, 
	mark size=0.7pt,
	point meta=explicit,
	each nth point=8
	] table [x=H2O,y=Z, meta=beta, col sep=comma]{../sources/Mascotte/Mascotte_scatter_midplane_betaMore095.csv};
\end{groupplot}
\node[anchor=north, yshift=-2ex, xshift=-7ex] (captiona) at (group c1r1.south west){a)};
\node[anchor=north, yshift=-2ex, xshift=-6ex] (captiona) at (group c2r1.south west){b)};
\end{tikzpicture}%
  \caption{Scattered data from five independent time steps: a) Temperature data as function of H$_2$O mass fraction in the $x$-$y$ plane colored by the vapor mole fraction $\beta$. b) Compressibility factor over the H$_2$O mass fraction colored by $\beta$. Black dots indicate a stable single-phase mixture or pure fluids.}
  \label{fig: Mascotte_scatter_midplane}
\end{figure}

By comparing Figs.~\ref{fig: Mascotte general flame features}~a, c and d it gets obvious that the region of phase separation is enclosed by the iso-lines for $f = 2\cdot10^{-4}$ and for $Z=0.99$. The phase separation is triggered by real-gas effects, moderate temperatures and the presence of water. In Fig~\ref{fig: Mascotte_scatter_midplane} these facts are further underlined by means of scatter data extracted at five independent time steps in the $x$-$y$ plane. Figure~\ref{fig: Mascotte_scatter_midplane}~a shows the temperature $T$ plotted other the H$_2$O mass fraction colored by vapor fraction $\beta$. The scatter data forms a kind of ribbon and the intersection separates the area of phase separation (left of the intersection) from the area of a stable single-phase mixture. A very similar shape results for the scattering of the compressibility factor $Z$, see Fig.~\ref{fig: Mascotte_scatter_midplane}~b. In this plot the determination of the phase separation region for $Z \approx 1$ can be retraced in detail which is especially true for the oxidizer-rich side.

\subsubsection{Comparison with the Experiment}
\begin{figure}[!h]
	\vspace*{12pt}
  \centering
   \input{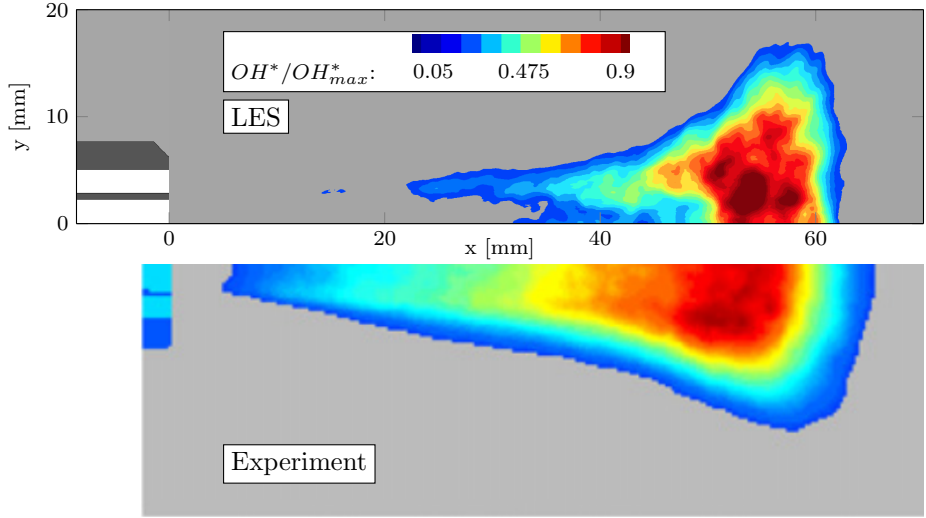}
   \vspace*{6pt}
  \caption{Line-of-sight integrated OH$^*$ emission in comparison with the averaged experimental OH$^*$ radiation of Singla \textit{et al.}~\cite{Singla2005}. (Experimental data reprinted from Singla \textit{et al.}~\cite{Singla2005}, Fig. 4(c), Copyright 2004 The Combustion Institute. Published by Elsevier Inc.).}
  \label{fig: Mascotte_OHstar_LoS}
\end{figure}

Experimental data is available from Singla \textit{et al.}~\cite{Singla2005} in terms of Abel-transformed and line-of-sight integrated time-averaged OH$^*$ emission images. In order to evaluate the numerical results we use the model of Fiala and Sattelmayer~\cite{Fiala2016} to extract the OH$^*$ emission from the simulation. Figure~\ref{fig: Mascotte_OHstar_LoS} compares the line-of-sight integrated OH$^*$ radiation with the computational results. The characteristic flame shape discussed above is also visible in the experiments, where a constant spreading angle is observed in the front part of the flame. Further downstream, expansion due to chemical conversion takes places and subsequently, an abrupt end of the reaction zone can be seen. The LES results properly reflect this behavior. However, Singla \textit{et al.}~\cite{Singla2005} did not publish the exact dimensions of the experimental field of view such that a more detailed validation is not possible. Nevertheless, the proportions of the flame in the LES in therms of shape and length indicate good agreement with the experiment.

\section{Conclusions}
A thermodynamic framework based on a cubic equation of state has been presented. The model takes into account both real-gas effects as well as multicomponent phase separation. The stability of the single-phase mixtures is evaluated based on the Gibbs free energy. If an unstable mixture is detected, a phase separation is computed using a flash algorithm yielding the appropriate phase composition and consistently calculated thermodynamic properties. The method has been applied to LOx/CH$_4$ and LOx/H$_2$ counterflow diffusion flames. Pressure and propellant inlet temperatures have been set according to well-known experiments featuring upper stage relevant conditions.

General real-gas effects of the one-dimensional flames have been discussed for both propellant combinations. The phase separation analysis showed unstable mixtures on both the oxidizer and fuel side of the flamelets at moderate temperatures below $T \lesssim 400$~K. Stronger effects are observed at the oxidizer side indicated by a minimum vapor mass fraction of approximately 0.97 compared to 0.998 at the fuel side in the case of LOx/CH$_4$. Similar results have been found for LOx/H$_2$.
 
By comparing the reference solution without phase separation to the result incorporating condensation effects, the following findings have been obtained: The difference between thermodynamic properties of the dense-gas and the multi-phase solution ranges up to almost 40\%. However, the general profiles of the properties in mixture fraction space are not significantly affected by the phase separation process and therefore the deviations are mostly caused by a shift to lower mixture fraction values. The shift occurs due to the higher temperatures in the two-phase solution. These higher temperatures result from the gas condensation and the maximum difference is approximately 40~K.
 
The results of the one-dimensional LOx/CH$_4$ flames have been tabulated and used for the Large-Eddy Simulation of a single-element test case. The spatial extent of the phase separation phenomena is mostly restricted to a region around the LOx core. This region is enclosed by the iso-line of almost pure oxygen and the transition to an ideal gas state. Line-of-sight integrated OH$^*$ emission images indicate that both flame length and shape are in good agreement with the experimental results. As expected from the one-dimensional analysis, the consideration of phase separation has no significant effect on the results.

		\subsection*{Acknowledgements}
		Financial support has been provided by the German Research Foundation (Deutsche Forschungsgemeinschaft -- DFG) in the framework of the Sonderforschungsbereich Transregio 40 and Munich Aerospace (www.munich-aerospace.de). The authors gratefully acknowledge the Gauss Centre for Supercomputing e.V. (www.gauss-centre.eu) for funding this project by providing computing time on the GCS Supercomputer SuperMUC at Leibniz Supercomputing Centre (LRZ, www.lrz.de).
		
		\vspace*{24pt}
		
		\bibliographystyle{unsrt}
		\bibliography{JPC2018}
		
	\end{document}